\documentclass[a4paper,12pt]{article}
\pdfoutput=1 

\usepackage{jheppub} 
\usepackage[export]{adjustbox}
\usepackage[utf8]{inputenc}

\usepackage{slashed}
\usepackage{bm}
\usepackage{multirow}

\usepackage{fancyvrb}
\usepackage{fvextra}
\usepackage{sidecap}

\usepackage{tikz}
\usepackage{tkz-euclide}
\usetikzlibrary{shapes.misc}
\usetikzlibrary{decorations.pathmorphing}	
\tikzset{
    v/.style={decorate, decoration={snake, segment length=3mm, amplitude=0.75mm}, draw},
    f/.style={draw=black, postaction={decorate},
        decoration={markings,mark=at position .6 with {\arrow[very thick]{latex}}}},
    fb/.style={dra w=black, postaction={decorate},
        decoration={markings,mark=at position .4 with {\arrowreversed[very thick]{latex}}}},
    fnar/.style={draw=black},
    g/.style={decorate, draw=black,
        decoration={coil,amplitude=3pt, segment length=3.5pt}},
    s/.style={dashed,draw=black, postaction={decorate},
        decoration={markings,mark=at position .55 with {\arrow[very thick]{latex}}}},
    sb/.style={dashed,draw=black, postaction={decorate},
        decoration={markings,mark=at position .55 with {\arrowreversed[draw=black,very thick]{latex}}}},
    snar/.style={dashed,draw=black,line width =1.25pt},
    cross/.style={cross out, draw=black, minimum size=2*(#1-\pgflinewidth), inner sep=0pt, outer sep=0pt},
cross/.default={3pt},
}

\graphicspath{{Figures/}} 

\newcommand{\al}[1]{\begin{align}\begin{aligned} #1 \end{aligned}\end{align}}

\newcommand{\A}[3]{A\hspace{-2pt}\left(#1, #2, #3 \right)}

\def\be{\begin{equation}}
\def\ee{\end{equation}}
\newcommand{\ba}{\begin{array}}
\newcommand{\ea}{\end{array}}

\usepackage[normalem]{ulem}

\definecolor{mtcolor}{rgb}{.8,.3,.1}

\title{\boldmath Probing light dark scalars with future experiments
}

\author[a]{Enrico Bertuzzo}
\author[b]{ and Marco Taoso}

\affiliation[a]{Instituto de F\'{i}sica, Universidade de S\~{a}o Paulo, C.P. 66.318, 05315-970 S\~{a}o Paulo, Brazil}
\affiliation[b]{I.N.F.N. sezione di Torino, via P. Giuria 1, I-10125 Torino, Italy}

\emailAdd{bertuzzo@if.usp.br}
\emailAdd{marco.taoso@to.infn.it}

\abstract{
We investigate a dark sector containing a pair of light non-degenerate scalar particles, with masses in the MeV-GeV range, coupled to the visibile sector through heavier mediators. The heaviest dark state is long-lived, and its decays offer new testable signals.
We analyze the prospects for detection with the proposed beam-dump facility SHiP, and the proposed LHC experiments FASER and MATHUSLA. Moreover, we consider bounds from the beam-dump experiment CHARM and from colliders (LEP, LHC and BaBar).
We present our results both in terms of an effective field theory, where the heavy mediators have been integrated out, and of a simplified model containing a vector boson mediator, which can be heavy $\gtrsim\mathcal{O}(1)$ TeV, or light $\mathcal{O}(10)$ GeV. 
We show that future experiments can test large portions of the parameter space currently unexplored, and that they are complementary to future High-Luminosity LHC searches.
}

\begin{document} 
\maketitle

\section{Introduction and framework}
\label{sec:intro}

Although elusive, it is a concrete possibility that dark sectors exist in nature. Their physics is interesting for various reasons: they may appear in connection to dark matter (DM)\,\cite{Battaglieri:2017aum}, they could produce new signals, such emerging jets at colliders (see e.g. hidden valley models\,\cite{Strassler:2006im,Strassler:2006ri,Han:2007ae}), and they might be connected to the solutions of
some open questions of particle physics, for instance the hierarchy problem (e.g. in Twin Higgs\,\cite{Chacko:2005pe} and relaxion\,\cite{Graham:2015cka} models). 
Several new experiments have been proposed in the recent years to target such sectors\,\cite{Anelli:2015pba,SHiP:2018yqc,Feng:2017uoz,Ariga:2019ufm,Curtin:2018mvb,Alpigiani:2020tva,Gligorov:2017nwh,Berlin:2018pwi,Akesson:2018vlm,Bauer:2019vqk}, and investigate possible extensions of the Standard Model (SM) of particle physics. 
Particular attention have been devoted to \textit{light} dark states, with masses in the MeV to GeV range. 
These scenarios can be explored with experiments both at the \textit{intensity} frontier\,\cite{Beacham:2019nyx} and at the \textit{energy} frontier, like the Large Hadron Collider (LHC), with a nice interplay between the two experimental programs. 

Dark sector physics can be broadly characterized in terms of the particle content in the dark sector itself, and the mediators that connect them to the SM. The interactions between the dark states and the SM can thus be written in terms of renormalizable or non-renormalizable portal operators.
The latter option is appropriate when the mediators are heavier than the energy scale involved in the process under consideration, like the production of dark sector particles in an experiment. In this case, the mediators can be integrated-out, and the physics is described in terms of contact non-renormalizable operators.
The advantage of this description in terms of an effective field theory (EFT) is its model-independence: it captures at the same time different microscopic ways in which the dark sector communicates with the SM.
When the mediators are light they can be produced on-shell in the experiments considered. 
A small drawback is that the description, 
now usually in terms of renormalizable operators, is more model dependent.
On the positive side, processes in which the mediator is produced on-shell can now be used to study the scenario, a possibility that was precluded in the EFT description.

In this paper we will consider both cases in the context of a dark sector containing non-degenerate scalar states.
As we are going to see, this framework has a rich phenomenology, since the decays of the heavier state open up new experimental venues for detection. A similar situation with non-degenerate fermion states has been considered in ref.\,\cite{Darme:2020ral}. The prototypical model that we have in mind is given by
a complex scalar singlet 
\be
\phi = \frac{\phi_1 + i \phi_2}{\sqrt{2}}\ ,
\ee
whose real and imaginary components are mass splitted. The mass splitting can be generated dynamically, and although in the following its origin will not be important, it is useful to consider a concrete example~\footnote{Another possibility is that the mass splitting is due to radiative corrections, in analogy to the case of the charged and neutral components of the pions.}. Take for instance the potential
\be\label{eq:potential}
V = - \mu_H^2 |H|^2 + \lambda_H |H|^4 + \mu_\phi^2 |\phi|^2 + \lambda_\phi |\phi|^4 + \lambda_{H \phi} |H|^2 |\phi|^2 + B \phi^2 + h.c. 
\ee
In the $B = 0$ limit the $U(1)$ symmetry of the potential forces $\phi_1$ and $\phi_2$ to be degenerate (unless the $U(1)$ is spontaneously broken, a situation that we will not consider in this paper). Once $B$ is turned on, it is easy to see that  a mass splitting is obtained. To quantify the mass difference we introduce the parameter 
\be
\delta = \frac{m_2 - m_1}{m_1}\ ,
\ee
where $m_2$ and $m_1$ are the masses of the heavier and lighter states, respectively. Generically it is technically natural to have a small $\delta$, since in the $\delta = 0$ limit a symmetry forcing the two components of the complex singlet to be degenerate is recovered. The phenomenology of scalar states interacting with the SM via the Higgs portal coupling $\lambda_{H\phi}$ have been studied thoroughly in the literature (see for instance\,\cite{Arcadi:2019lka} for a recent review). In the following we will thus assume that the Higgs-portal coupling $\lambda_{H\phi}$ in eq.\,\eqref{eq:potential} is negligible, with the dark sector-SM interactions mediated by some new particle. As an example, we can imagine a vector boson $Z^{\prime}$ interacting via the Lagrangian
\be\label{eq:heavy_Zp}
\mathcal{L}_{int} = Z_\mu^{\prime} \left( g_{\phi} J_\phi^\mu+ \sum_{f_L} g_L^f \bar{f}_L \gamma^\mu  f_L + \sum_{f_R} g_R^f \bar{f}_R \gamma^\mu f_R \right)\ ,
\ee
where $f_L$ and $f_R$ are the left-handed and right-handed SM fermions, and the dark current is given by 
\be\label{eq:dark_current}
J_\phi^\mu =i [(\partial^\mu \phi^\dag) \phi - \phi^\dag (\partial^\mu \phi)]  =  (\partial^\mu \phi_2) \phi_1 - \phi_2 (\partial^\mu \phi_1)\ .
\ee 
Another possibility is to introduce vector-like fermions $\Psi_{L,R}^f$ interacting with the SM via
\be\label{eq:heavy_fermions}
\mathcal{L}_{int} = \sum_{f_L} y_L^f \phi\,  \bar{\Psi}_R^f f_L + \sum_{f_R} y_R^f \bar{\Psi}_L^f f_R + h.c.
\ee
Our focus here is on mediators with masses above few tens of GeV. The phenomenology of a light (below the GeV) mediator is different, and it has been studied elsewhere, see e.g.\,\cite{Curtin:2018mvb,Feng:2017uoz,SHiP:2020noy,Berlin:2020uwy,Berlin:2018jbm,Berlin:2018bsc,Berlin:2018pwi}.
Assuming the mediators to be heavy allows to give a unified treatment of the phenomenology at low energy.
Indeed, integrating out the vector in eq.\,\eqref{eq:heavy_Zp} or the fermions in eq.\,\eqref{eq:heavy_fermions} we obtain effective operators of the form\,\cite{Craig:2019wmo}:
\be\label{eq:EFT}
\mathcal{L}_{EFT} = \frac{J_\phi^\mu}{\Lambda^2} \left( \sum_{f_L} c_{f_L} \bar{f}_L \gamma_\mu f_L + \sum_{f_R} c_{f_R} \bar{f}_R \gamma_\mu f_R \right)+ \dots
\ee
where the dots represent other operators that are generated.~\footnote{In the case of heavy fermions, gauge invariance allows for Yukawa interactions of the type $g \bar{\Psi}^f_L H \Psi_R^f$ in addition to the operators in eq.\,\eqref{eq:heavy_fermions}. These interactions would generate additional dimension-6 operators of the form $\phi^\dag \phi \bar{f}_L H f_R$ which may contribute to the phenomenology of the model. In the following we will suppose the Wilson coefficient of such operators to be suppressed. This can be achieved for instance requiring Minimal Flavor Violation, in which case the Wilson coefficient is proportional to the fermion mass and is typically suppressed.} Notice that we write the operators in a $SU(3)_c \times SU(2)_L \times U(1)_Y$ invariant manner. 
In terms of vector and axial currents they are defined as
\be\label{eq:EFT2}
{\cal L}_{\rm EFT} = \frac{J_\phi^\mu}{\Lambda^2} \sum_f \bar{f} \gamma_\mu (c_{V_f} + c_{A_f} \gamma_5 ) f \ ,
\ee
where the sum is extended over all SM fermions. 
The Wilson coefficients  are simply
\be\label{eq:vector_axial_coupls}
c_{V_f} = \frac{c_{f_L} + c_{f_R}}{2} \ , ~~~~~ c_{A_f} = \frac{c_{f_R} - c_{f_L}}{2}\ .
\ee
Analogously, the vector and axial couplings of the $Z^{\prime}$ vector with the SM fermions are
\be\label{eq:vector_axial_coupls_Zp}
g_{V_f} = \frac{g_{L}^f + g_{R}^f}{2} \ , ~~~~~ g_{A_f} = \frac{g_{R}^f - g_{L}^f}{2}\ .
\ee
For definitiveness, in this paper we will consider only the case $c_{f_L} = c_{f_R}$ for the EFT operators, and $g_L^f = g_R^f$ for the couplings of the $Z'$ boson. This means that the SM fermions interact with the dark scalars only through vector currents, 
with the exception of the neutrinos, which couple through the usual V-A current.
We leave a detailed analysis of the phenomenology of axial interactions to future work. 
Notice also that this choice allows us to neglect SM radiative corrections that could otherwise be important. 
In fact, it has been shown that, for vector currents, the renormalization group evolution of the effective operators
from the scale $\Lambda$ to the scale of low energy experiments is small\,\cite{Hill:2011be,Frandsen:2012db,Vecchi:2013iza,Crivellin:2014qxa,DEramo:2014nmf,DEramo:2016gos,Brod:2017bsw,Brod:2018ust,Belyaev:2018pqr,Beauchesne:2018myj,Chao:2016lqd,Arteaga:2018cmw}.
Therefore, we will neglect such effect.

The operators introduced above control the production of the dark states from the scattering of SM particles, and the decay of $\phi_2$ into $\phi_1$ and SM states.
The latter process is suppressed in the limit of a small mass splitting $\delta$, and/or feeble interactions among the SM and the dark sector.
Interestingly, this opens the possibility that $\phi_2$ is long-lived, and travels macroscopic distances before decaying.
Such scenario can be tested in experiments where the dark states are produced in high-intensity or high-energy facilities, and then show up in a far-placed detector through the decays of $\phi_2.$
The aim of this paper is to explore the sensitivity of such experimental facilities for the scenario presented above.
Concretely, we will consider fixed target experiments, and proposals for detectors to be placed near the LHC interactions points. 
For the first class of experiments, we will compute the constraints from searches of the CHARM experiment\,\cite{Bergsma:1983rt}, and the projected sensitivity for the proposed SHiP facility\,\cite{Anelli:2015pba,SHiP:2018yqc}.
We will then focus on the LHC experiments FASER\,\cite{Feng:2017uoz,Ariga:2019ufm} and MATHUSLA\,\cite{Curtin:2018mvb,Alpigiani:2020tva}. 
We will also compare the sensitivity regions of these experiments with the collider bounds from searches at LEP, BaBar, and LHC (namely LHCb, ATLAS and CMS).

In our work we assume that $\phi_1$ is stable or very long-lived, such that once produced, it decays far outside the detectors that we are considering.
Although not essential, it is interesting to conceive the possibility that it is a DM candidate. We will briefly comment about this option later.

Before entering into the details of the calculations, let us explain the structure of the paper.
In section\,\ref{sec:CHARM_SHIP} and\,\ref{sec:FASER_MATHUSLA} we describe how the the constraints and future sensitivities are computed in fixed target experiments and LHC experiments.
In section\,\ref{sec:collider_searches} we present collider bounds which do not rely on the inelastic nature of the dark sector.
More specifically we will discuss constraints from searches of missing energy events at LEP, LHC and BaBar, as well as limits from the invisible decays of heavy quarkonia states. 
In section\,\ref{sec:bounds_EFT} we present our results in the context of the EFT operators in eq.\,\eqref{eq:EFT}. 
As mentioned before, we will consider mediators with masses above few tens of GeV. If light enough, these particles can be produced on-shell at LHC. Motivated by this consideration, in section\,\ref{sec:on_shell_med} we will move to the case of the $Z^{\prime}$ model in eq.\,\eqref{eq:heavy_Zp}. 
This will allow also a more precise comparison with current LHC searches. We will consider the case of a heavy $Z^{\prime},$ with a mass above few TeV, and the benchmark case of a dark photon with a mass of 40 GeV.
In section\,\ref{sec:cosmo} we present cosmological and astrophysical constraints. Part of this discussion applies only when $\phi_1$ is the dominant component of DM.
Finally, we conclude in section\,\ref{sec:conclusions}. We also add two appendices. In appendix\,\ref{app:useful_eqs} we collect the decay widths used in the EFT analysis, explaining in detail how they are obtained. In appendix\,\ref{app:Zp_decay} we instead present the decay widths of the $Z'$ boson, used for the computations in section\,\ref{sec:on_shell_med}.

\section{Overview of experimental bounds and forecasts}\label{sec:exp}
\begin{figure}[tb]
\begin{center}
$\includegraphics[width=.49\textwidth]{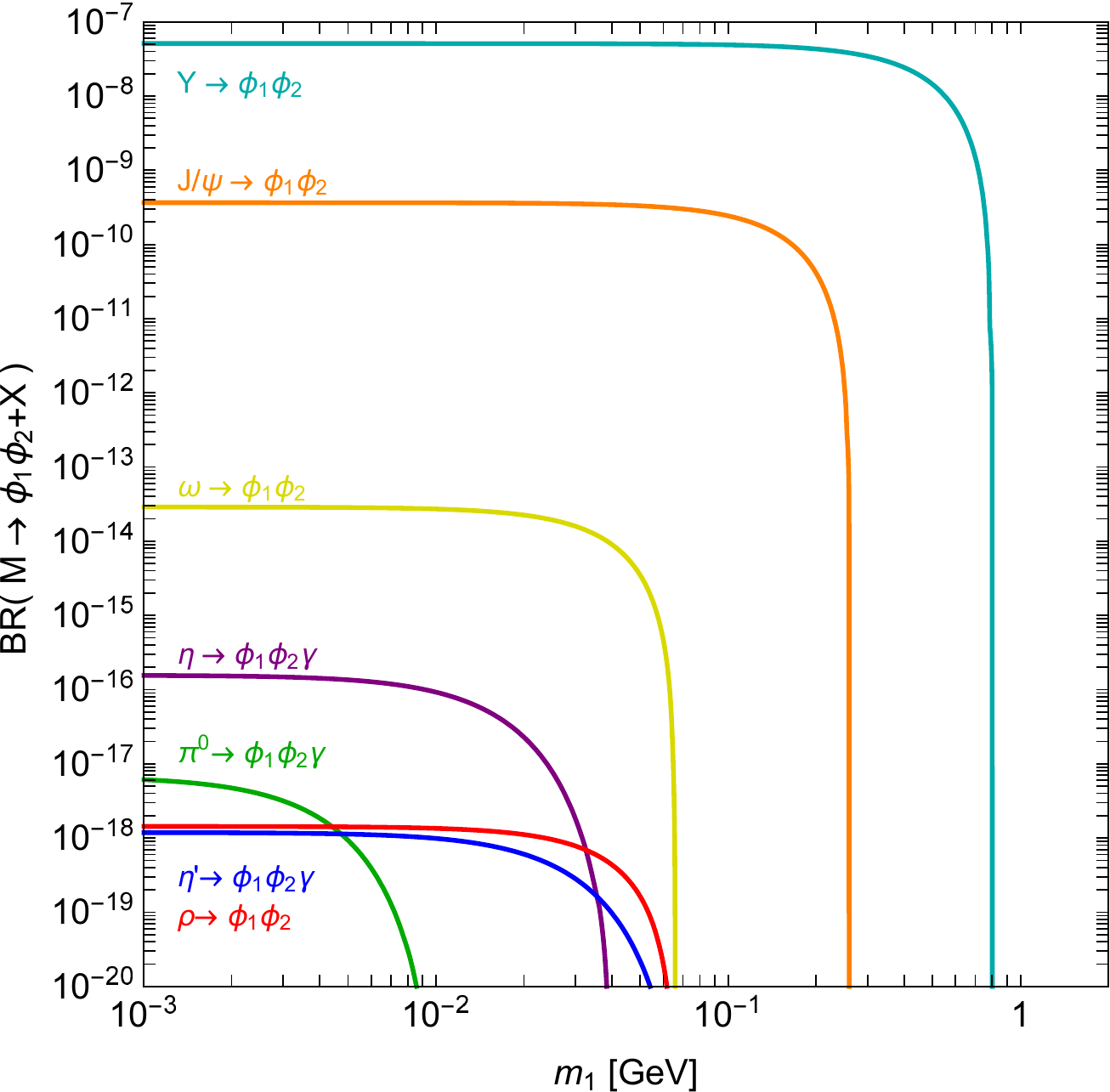}
\quad
\includegraphics[width=.49\textwidth]{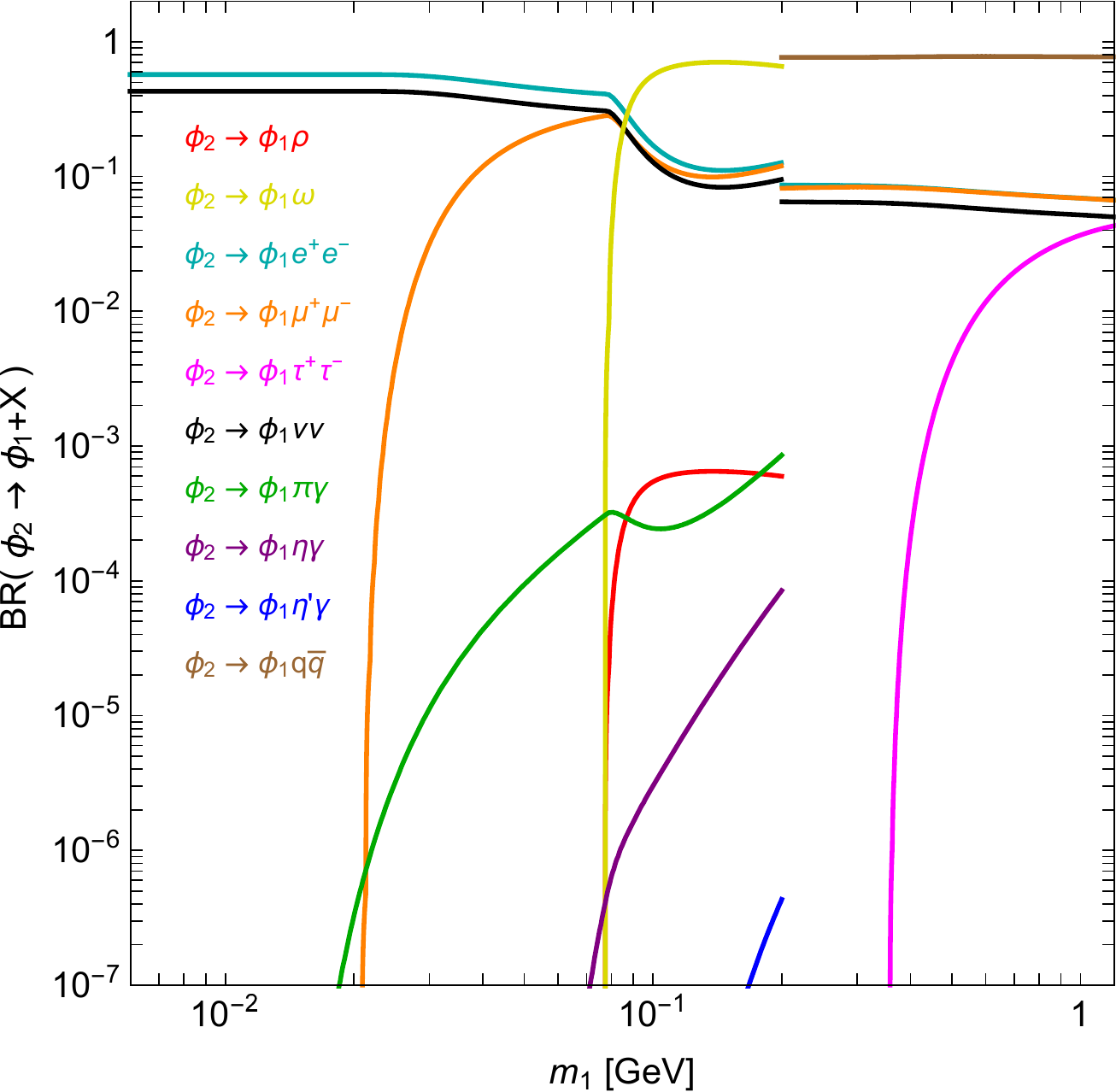}$
\caption{\label{fig:Decays} \em Left panel: branching ratio of meson decays into dark states. Right panel: branching ratio of $\phi_2$ decays into $\phi_1$ and SM states. In both panels we take $c_{f_L} = c_{f_R} = 1$, $\delta=10$ and $\Lambda=1$ TeV. In the right panel the threshold at $m_1\simeq$ 0.2 GeV denotes the transition of the treatment of $\phi_2$ hadronic decays from chiral perturbation theory to perturbative QCD. See appendix\,\ref{app:useful_eqs} for details.}
\end{center}
\end{figure}
 In the following two sections we examine current and future experiments that can test the interactions in eq.\,\eqref{eq:EFT} via $\phi_2$ decays, i.e. in which the inelastic nature of the dark sector is instrumental in probing the parameter space. 
We classify them according to their center-of-mass energy: $\sqrt{s} \simeq 28 $ GeV (section\,\ref{sec:CHARM_SHIP}) and $\sqrt{s} = 14$ TeV (section\,\ref{sec:FASER_MATHUSLA}). The physical processes of interest can be summarized as
\al{
p + \mathrm{target} & \to \phi_1 + \phi_2  & ~~&{\rm (production)} \\
\phi_2 & \to \phi_1 + \mathrm{SM} & ~~&{\rm (decay)}
}
If the decay of the heaviest state $\phi_2$ occurs inside the decay volume of the experiment, a signal event may be detected. 
$\phi_1 \phi_2$ pairs can be produced directly from parton collisions, and in the decays of the mesons generated by the proton interactions~\footnote{A third production mechanism, which we do not consider, is the proton bremsstrahlung: $p\,A\rightarrow p\,A\,\phi_1\phi_2,$ where $A$ is a nucleus in the target. This process might allow to extend our sensitivities to larger dark scalar masses. See e.g. the prospects for FASER\,\cite{Feng:2017uoz,Ariga:2019ufm} and SHiP\,\cite{SHiP:2020noy} for scenarios including a light dark photon.
}.

\subsection{Experiments at $\sqrt{s} \simeq 28$ GeV: CHARM and SHiP}\label{sec:CHARM_SHIP}

The CHARM\,\cite{Bergsma:1983rt} and the proposed SHiP\,\cite{Anelli:2015pba,SHiP:2018yqc} experiments exploit a $400$ GeV proton beam impinging on a fixed target, i.e. $\sqrt{s} \simeq 28$ GeV.
For these experimental facilities, we compute the fluxes of dark scalars generated by meson decays.
We do not consider the production of $\phi_1\, \phi_2$ pairs from parton level processes
for reasons that will be explained in sec.\,\ref{sec:FASER_MATHUSLA}. We include in our analysis light neutral pseudo-scalar mesons ($\pi^0,$ $\eta$ and $\eta^{\prime}$), neutral vector mesons ($\rho$ and $\omega$) and the charmonium $J/\psi$ and the bottomonium $\Upsilon(1S)$ resonances. 

The light pseudo-scalar and vector mesons are more abundantly produced by the proton interactions than the heavier ones.
Nevertheless, the latter tend to have larger branching ratios into dark states, that may compensate for the smaller production cross sections. We will come back to this point later. Furthermore, heavy mesons are obviously relevant for heavy enough dark states, whose production from light mesons is kinematically forbidden. 
We show in the left panel of fig.\,\ref{fig:Decays} the branching ratios for the decays of the mesons into $\phi_1\,\phi_2$ states, for the following benchmark case: $\Lambda = 1$ TeV, $\delta = 10.$ In this plot, and in the rest of the paper for the discussion of the EFT operators in eq.\,\eqref{eq:EFT}, we adopt a democratic vector coupling to all the SM fermions, specifically $c_{f_L}=c_{f_R}=1$ for all the SM fermions (of course $c_{f_R}=0$ for the neutrinos). Detailed formulas for the decay widths can be found in appendix\,\ref{app:useful_eqs}. 
For vector interactions, the dominant  decay mode of the light pseudo-scalar mesons into dark sector particles is via the three-body process $M \to \phi_1 \phi_2 \gamma$.
Notice also that the branching ratio for $\rho \to \phi_1 \phi_2$ is suppressed with respect to the one for the analogous processes involving the other vector mesons ($\omega$, $J/\psi$ and $\Upsilon$). This is because the effective coupling of the $\rho$ meson to the dark current is suppressed, for the choice of couplings above (see eq.\,\eqref{eq:f_rho_omega}).
As mentioned before, $J/\psi$ and $\Upsilon$ have larger branching fraction into the dark states than the lighter mesons.
 
Our goal is to compute the number of signal events in a detector. This can be done using: 
\be\label{eq:total_phi2_mes}
N_{\rm sig} = \sum_M N^{\phi_2}_{{\it{M}}} \,f^{\rm dec}_M \, f^{\rm rec}_{M}\ ,
\ee
where $N^{\phi_2}_{{\it{M}}}$ is the number of $\phi_2$ particles produced in the decays of the meson $M,$ $f^{\rm dec}_M$ is the fraction of events in which $\phi_2$ decays inside the detector, and $f^{\rm rec}_{M}$ is the efficiency for the reconstruction of the signal event in the detector. We are going to discuss how to compute each term.\\

\begin{table}[tb]
\begin{center}
\begin{tabular}{c|cccccccc}
\hline
& $\pi^0$ & $\eta$ & $\eta^{\prime}$ & $\rho$ & $\omega$& $J/\psi$&$\Upsilon$\\
\hline\hline
 $N_{{\it{m}}}$ (CHARM, SHiP) & 4.0 & 0.44 & 0.046 & 0.51 &0.45 & $6.25\times 10^{-6}$&$2.25\times 10^{-9}$\\  
 $N_{{\it{m}}}$ (LHC experiments) & 38 & 4.16 & 0.45 & 4.6 & 4.3 & $7.9 \times 10^{-4}$ & $8.4 \times 10^{-6}$
\end{tabular}
\end{center}\vspace{-0.35cm}
\caption{\em Average number of mesons produced per proton interaction for fixed target experiments (CHARM, SHiP) and for experiments at the LHC (FASER, MATHUSLA). See the text for details on how these multiplicities were computed.}
\label{tab:multiplicities}
\end{table}
\noindent\textbf{Production from meson decays.} The number of dark particles produced in the decays of the mesons $M = \left\{ \pi^0, \eta, \eta', \rho, \omega, J/\psi, \Upsilon\right\}$ can be computed according to
\be
N^{\phi_2}_{{\it{M}}} = {N_{\rm POT}}\,N_{{\it{M}}}\, \rm{BR}({\it{M}}\rightarrow\phi_1\,\phi_2+X),
\label{eq:prodMes} 
\ee
where $N_{\rm POT}$ is the number of protons on target collected (for CHARM) or expected (for SHiP) by the experiment, $N_M$ the average number of mesons produced per proton interaction, and $\rm{BR}({\it{M}}\rightarrow\phi_1\,\phi_2+X)$ is the branching ratio of the meson $M$ into the dark states. The number of protons on target for the CHARM and SHiP experiments\,\cite{Bergsma:1985is,Anelli:2015pba,SHiP:2018yqc} are summarized in table\,\ref{tab:exp_specs}. 

The average number of mesons produced per proton interaction, $N_M$, is computed in the following way. We simulate $pp$ collisions using two different softwares: {\tt EPOS-LHC}\,\cite{Pierog:2013ria} (as found in the {\tt CRMC}  package\,\cite{crmc}) for light mesons, and {\tt{PYTHIA8}}\,\cite{Sjostrand:2007gs} for the $J/\psi$ and $\Upsilon$ mesons.
Since {\tt EPOS-LHC} has been tuned to correctly reproduce mesons multiplicities and energy distributions of LHC data, it is particularly useful for our purposes. 
For the heavy mesons we use {\tt{PYTHIA8}}, which allows to switch on only the charmonium or bottomonium production processes~\footnote{More specifically, we use the flags ``{\tt Charmonium:all}'' and ``{\tt Bottomonium:all}'' to generate charmonium and bottomonium events respectively.}.
We find the total production cross-sections $\sigma_{J/\psi} \simeq 250$ nb and $\sigma_{\bar{b}b} \simeq 1.3$ nb, in reasonable agreement with the literature\,\cite{Abt:2005qr,Lourenco:2006vw,Patrignani:2016xqp}.
From the results of our simulations, and using a total proton-proton cross section $\sigma_{pp} \simeq 40$ mb\,\cite{CERN-SHiP-NOTE-2015-009}, we 
compute the average number of mesons produced per proton interaction, reported in table\,\ref{tab:multiplicities}.
For light mesons, we have checked that similar multiplicities are obtained using {\tt{PYTHIA8}}.
A comparison between the output of {\tt{PYTHIA8}} and experimental data\,\cite{AguilarBenitez:1991yy} for the production of $\pi^0$ and $\eta$ mesons can be found in\,\cite{Dobrich:2019dxc}~\footnote{The SHiP collaboration have compared the samples of $\pi^0$ and $\eta$ mesons obtained with {\tt{PYTHIA8}} with those from their software, based on {\tt{GEANT4}}\,\cite{Allison:2016lfl}, which takes into account secondary interactions of hadrons in the target\,\cite{SHiP:2020noy}. For the scenario that they are considering, they find that cascades affect the signal rate by 15-40\% for $\pi^0$ decays, while the impact is negligible for the case of $\eta$ mesons.}.

\begin{figure}[tb]
\begin{center}
$\includegraphics[width=.49\textwidth]{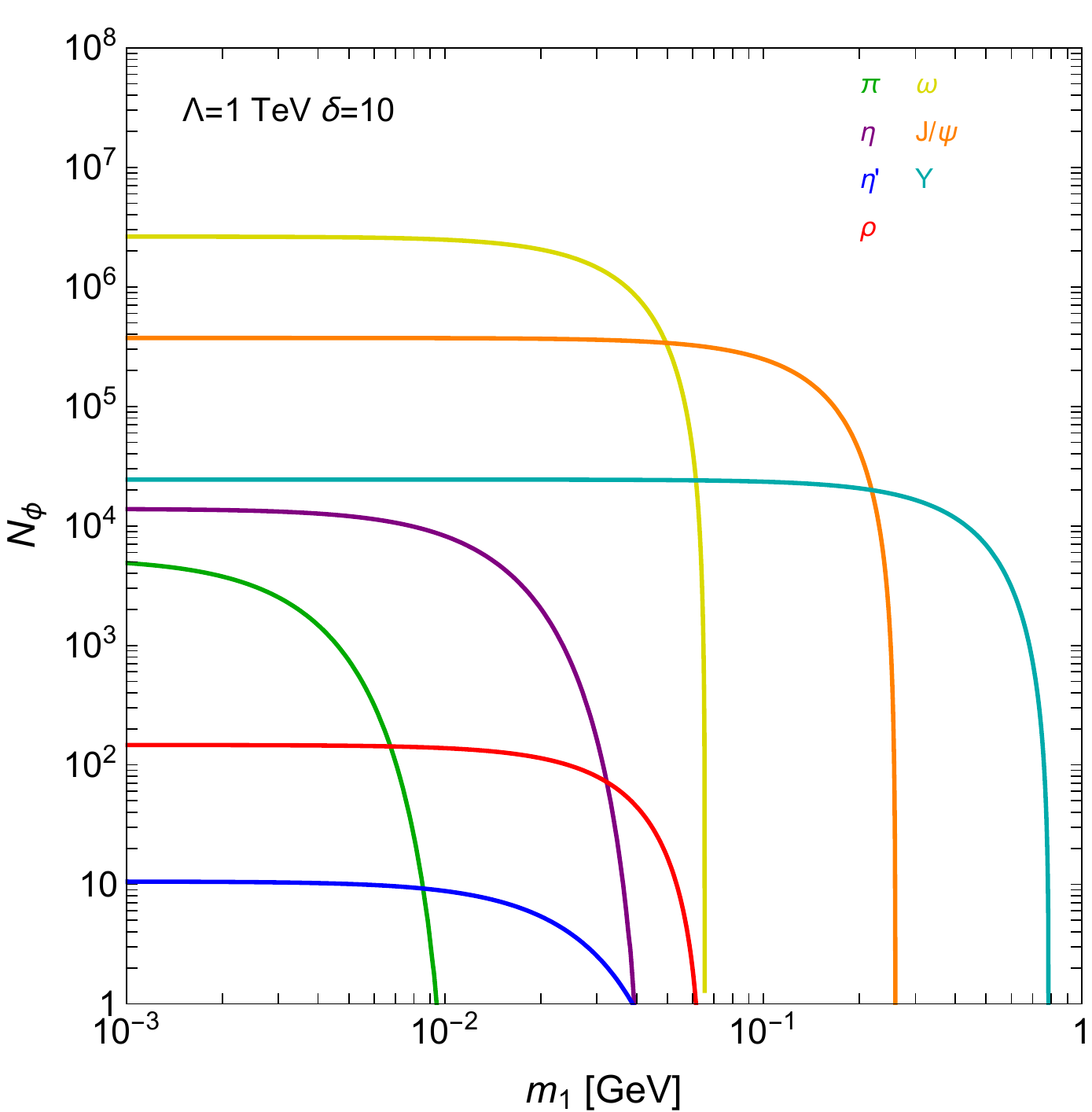}
\quad
\includegraphics[width=.49\textwidth]{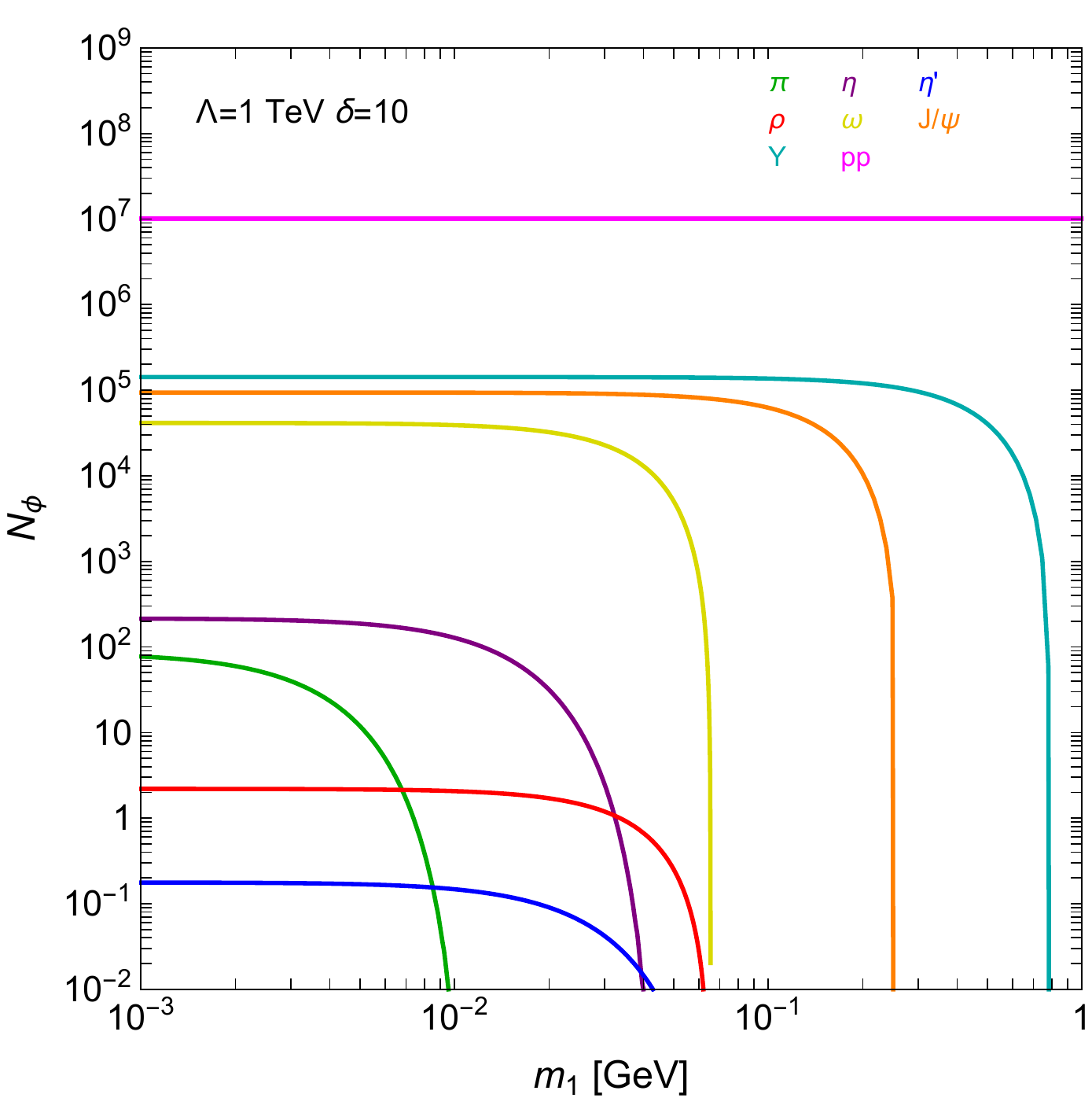}$
\caption{\label{fig:Nmesons} \em Number of $\phi_2$ particles produced in proton collisions in two different experiments: SHiP (left panel) and MATHUSLA (right panel). In the case of SHiP we consider only the production via mesons decays. For MATHUSLA we include also direct parton production (pp). In both panels we fix $c_{f_L} = c_{f_R} = 1$, $\delta=10$ and $\Lambda=1$ TeV.
 }
\end{center}
\end{figure}

Finally, the branching ratios $\rm{BR}({\it{M}}\rightarrow\phi_1\,\phi_2+X)$ are obtained using the equations in appendix\,\ref{app:useful_eqs}.
They are shown in the left panel of fig.\,\ref{fig:Decays} for the same benchmark case discussed before. In fig.\,\ref{fig:Nmesons} (left panel) we show the expected number of $\phi_2$ particles at SHiP. 
The production via mesons decays is dominated by the $\omega$ contribution for small dark scalar masses, and by the $J/\psi$ and $\Upsilon$ decays for larger ones.
Notice that the contributions from $J/\psi$ and $\Upsilon$ exceed those from the light mesons, except for the $\omega.$ As anticipated, this is due to their larger branching ratio into dark states.

Let us also mention that the relative contributions are different when the mediator of the interaction between the dark and visible sectors is light so that it can be produced on-shell, see e.g.\,\cite{SHiP:2020noy,Berlin:2018pwi}.\\

\noindent{\bf Dark state decays.} The fraction of $\phi_2$ states which decay inside the detector can be computed as:
\be\label{eq:Ndec}
f^{\rm dec}_M = f^{\rm geom}_M\, \left\langle e^{-L/L_{\phi_2}} - e^{-(L+L_{\rm dec})/L_{\phi_2}} \right\rangle\ .
\ee
The first factor is a geometrical cut and represents the fraction of $\phi_2$ particles whose trajectories intersect the detector. The second term takes into account the probability that those particles decay inside the detector.
It is computed in terms of the distance between the interaction point and the detector ($L$), the detector size ($L_{\rm dec}$) and the decay length of $\phi_2$ in the laborarory (LAB) frame, given by $L_{\phi_2}= c\,\tau_{\phi_2}\gamma_{\phi_2}\beta_{\phi_2}$. Here $\tau_{\phi_2}$ is the decay time of $\phi_2,$ while $\beta_{\phi_2}$ 
and $\gamma_{\phi_2}$ are respectively its speed in units of speed of light ($c$), and its Lorentz factor, in the LAB frame.

The decay time $\tau_{\phi_2}$ is obtained using the equations of appendix\,\ref{app:useful_eqs}. The additional quantities are obtained as follows.
We consider the samples of mesons obtained with {\tt EPOS-LHC} and {\tt{PYTHIA8}}. For each event, we simulate the $M \to \phi_1 \phi_2 +X$ decay in the rest frame of the meson, and then compute the 4-momentum of $\phi_2$ in the LAB frame performing the appropriate Lorentz boost. From this, one can compute the $\beta_{\phi_2}$ and $\gamma_{\phi_2}$ factors. Then, we select only those events whose trajectories intersect the detector, which allows us to compute $f^{\rm geom}_M$.
For this purpose, we approximate the detectors as cylinders, and we impose a cut on $\theta_{\phi_2}$, the angle between the direction of flight of $\phi_2$ and the beam axis.
In the case of SHiP the area of the cylinder is taken to be $A_{\rm dec}=5\times10\,\rm{m}^2$\,\cite{Anelli:2015pba,SHiP:2018yqc}. The maximum opening angle $\theta_{\phi_2}$ is thus $\tan\left( \theta_{\rm dec} \right) = \sqrt{A_{\rm dec}/\pi}/(L+L_{\rm dec})$.
For CHARM we follow ref.\,\cite{Dobrich:2015jyk} to take into account the fact that the detector is off-axis. We summarize all the relevant quantities in table\,\ref{tab:exp_specs}.
The second term in eq.\,\eqref{eq:Ndec} is computed averaging ($\langle \cdot \rangle$) the probabilities over all the events in the direction of the detector.\\

\noindent{\bf Reconstruction of the events in the detector.} To mimic event selection cuts of experimental analysis, we impose a requirement on the energy of the visible particles produced in $\phi_2$ decays.
More specifically we adopt the following simplified strategy. 
We focus on the dominant decays $\phi_2\rightarrow\phi_1+V$ (where $V$ is $\rho$ or $\omega$) and $\phi_2\rightarrow\phi_1+\it{l}\bar{\it{l}}$ (where $\it{l}$ is a charged lepton), see fig.\,\ref{fig:Decays}.\,\footnote{The cut around $m_1 \simeq 0.2$ GeV is due to a change of description of the $\phi_2$ decays between chiral perturbation theory and perturbative QCD. See appendix\,\ref{app:useful_eqs} for more details.
}
From the sample of $\phi_2$ events produced in the decay of a given meson $M$, we compute the energy of the meson $V$ or of one of the charged leptons $\it{l}$ produced in the decay. Then, we select the events where the energy of these \emph{visible} particles is larger than a certain threshold $E_{\rm cut}$ to be specified below. 
We compute the fraction of events which satisfy this energy requirement, $f^{\rm rec, i}_M,$ for each of the $\phi_2$ decay channel $i$ under consideration.
Finally, we define the efficiency of this selection cut as
\begin{align}
f^{\rm rec}_M =\sum_i {\rm BR}(\phi_2 \to \phi_1 + {\rm X}_i) \, f^{\rm rec, i}_M \,\epsilon_i\ ,
\label{eq:effE} 
\end{align}
where ${\rm BR}(\phi_2 \to \phi_1 + {\rm X}_i)$ is the branching ration for the decay of $\phi_2$ in the channel $i$.
In the equation above, we have also introduced an efficiency $\epsilon_i$ for the reconstruction of the visible states in the decay. 
For CHARM, we recast a search of heavy neutrinos decaying into light neutrinos and an electrons or muons pair\,\cite{Bergsma:1985is}. We impose a cut $E_{\rm cut} = 2$ GeV and we adopt the efficiencies for the the electron and muon channels in table\,\ref{tab:exp_specs}. 
For SHiP, instead, we simply assume $E_{\rm cut} = 2$ GeV and a $100\%$ efficiency in all channels. The branching ratio of $\phi_2$ in the relevant channels can be computed using the formulas in appendix\,\ref{app:useful_eqs}. 
\begin{table}[tb]
\begin{center}
\begin{tabular}{c|c|c|c|c|c}
& $N_{\rm POT}/{\cal L}$ & $\theta_{\phi_2}$ [mrad] & $L$ [m] & $L_{\rm dec}$ [m] & $\epsilon$ \\ 
\hline
CHARM & \multirow{2}{*}{$2.4 \times 10^{18}$}& \multirow{2}{*}{$6.8 \leq \theta_{\phi_2} \leq 12.6$} & \multirow{2}{*}{480} & \multirow{2}{*}{35} & $\epsilon_e = 0.65$ * \\
(mes) & & & &  & $\epsilon_\mu = 0.75$ * \\
\hline
SHiP & \multirow{2}{*}{$ 2 \times 10^{20}$} & \multirow{2}{*}{$|\theta_{\phi_2}| \leq 36.9$} & \multirow{2}{*}{58} & \multirow{2}{*}{50} & \multirow{2}{*}{$\epsilon_i = 1$}  \\
(mes)  & & & &  & \\
\hline
FASER  &  \multirow{2}{*}{$3\, {\rm ab}^{-1}$} &  \multirow{2}{*}{$|\theta_{\phi_2}| \leq 2.1$} &  \multirow{2}{*}{480} &  \multirow{2}{*}{5} &  \multirow{2}{*}{$\epsilon_i = 1$} \\
(mes)  & & & &  & \\
\hline
MATHUSLA & \multirow{2}{*}{$3\, {\rm ab}^{-1}$} & \multicolumn{3}{c|}{\multirow{2}{*}{eq.\,\eqref{eq:Mathusla_geom}}} & \multirow{2}{*}{$\epsilon_i =1$} \\
(mes + pp)  & &\multicolumn{3}{c|}{} & \\ 
\end{tabular}
\end{center}\vspace{-0.35cm}
\caption{\label{tab:exp_specs} \em Specifications used for each experiment to compute the total number of $\phi_2$ signals in the detector according to eq.\,\eqref{eq:total_phi2_mes}: $N_{\rm POT}$ is the number of proton on target (for CHARM and SHiP), ${\cal L}$ is the total luminosity of the experiment (for FASER and MATHUSLA), $\theta_{\phi_2}$ is the angle between the direction of flight of $\phi_2$ and the beam axis, $L$ is the distance between the interaction point and the detector, $L_{\rm dec}$ is the detector length and $\epsilon$ is the efficiency for detection of the $\phi_2$ decay products. For each experiment we also specify the production mechanism for $\phi_2$ that we have considered (mes = meson decays, pp = direct parton production). (*) In the case of the CHARM experiment the efficiencies must be multiplied by a factor $0.095$ to take into account that the detector covers only $9.5\%$ of the circular corona defined by the angular cut. }
\end{table}
\subsection{Experiments at $\sqrt{s} = 14$ TeV: FASER and Mathusla}\label{sec:FASER_MATHUSLA}
\begin{figure}[tb]
\begin{center}
$
\includegraphics[width=.49\textwidth]{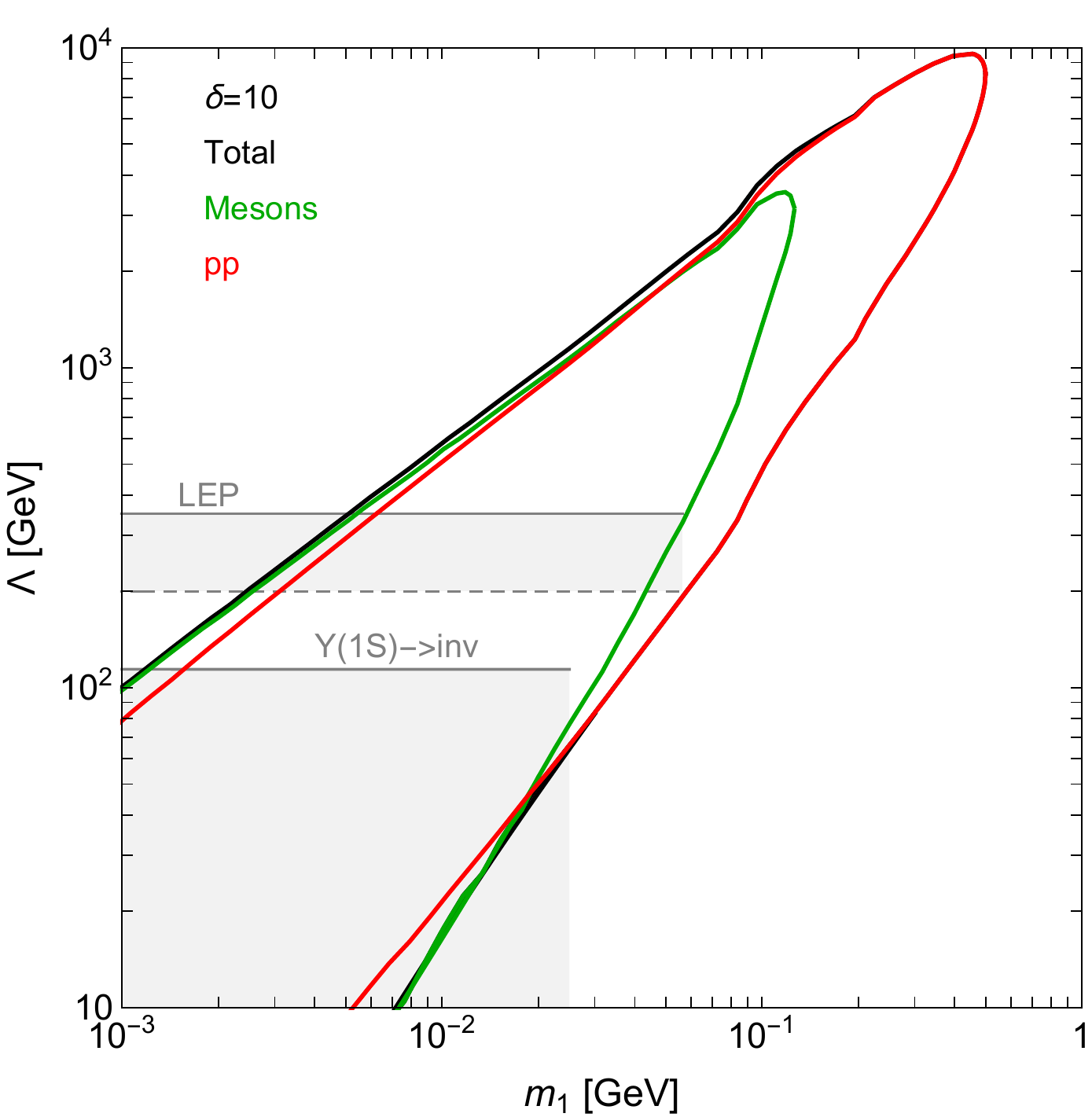}
\quad
\includegraphics[width=.49\textwidth]{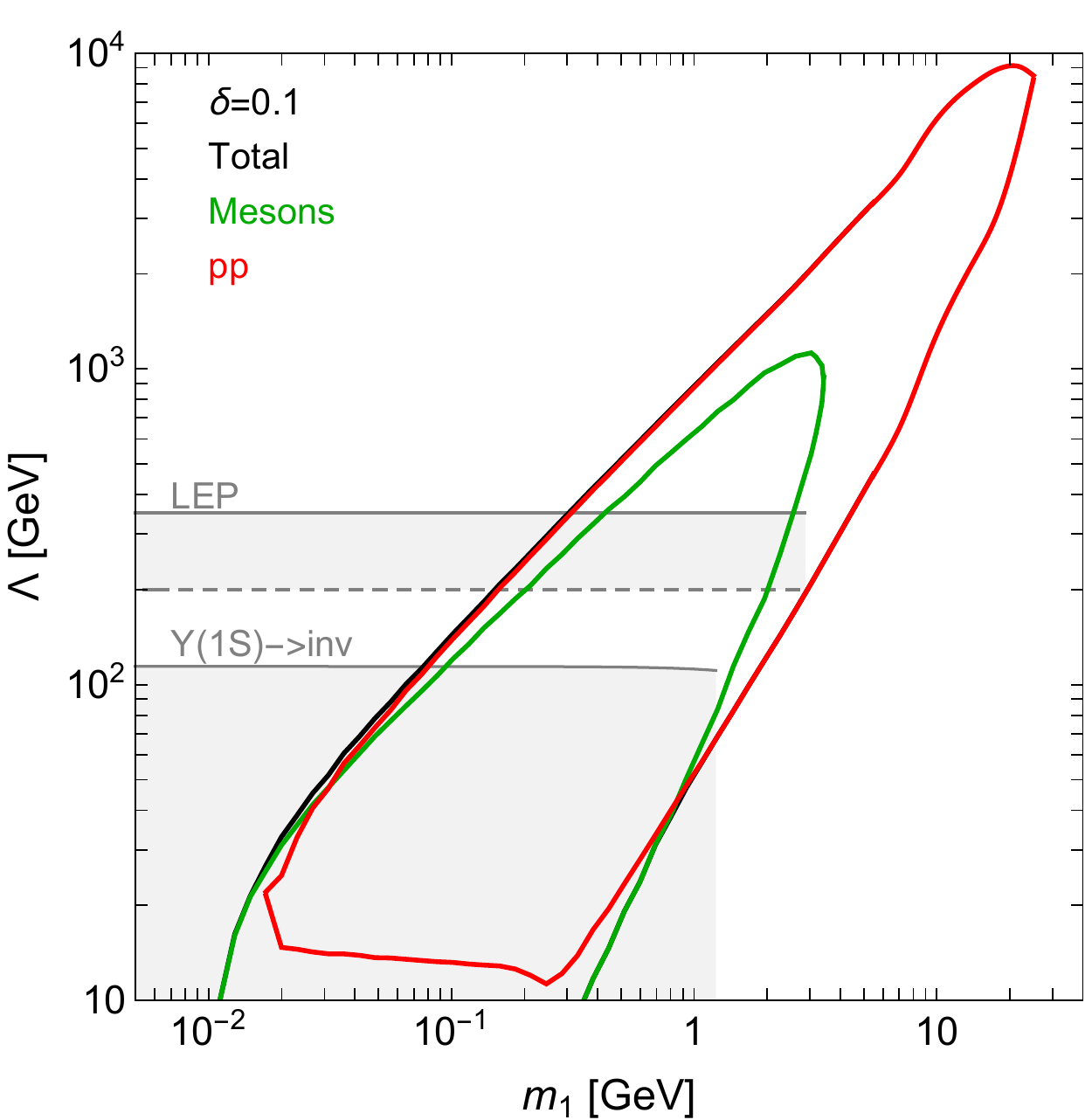}
$
\caption{\label{fig:Mathusla} \em 90\% CL projected sensitivity of the MATHUSLA experiment to the scale $\Lambda$ of EFT operators in eq.\,\eqref{eq:EFT}, as a function
of the mass $m_1$ of the lightest dark scalar. Left (right) panel is for a mass splitting of $\delta = 10$ ($\delta = 0.1$).
We take $c_{f_L} = c_{f_R} = 1$ in both panels.
Green (red) lines are for the production from meson decays (direct parton production) only.
The black lines include both production processes.
Gray regions are excluded by the constraints discussed in sec.\,\ref{sec:collider_searches}.
}
\end{center}
\end{figure}

We consider two proposed experiments that aim to leverage the proton collisions occurring at $\sqrt{s} = 14$ TeV at LHC: FASER\,\cite{Feng:2017uoz,Ariga:2019ufm} and MATHUSLA\,\cite{Curtin:2018mvb,Alpigiani:2020tva}. The FASER detector is planned to be placed downstream of the ATLAS experiment along the beam axis.
MATHUSLA instead will be located on the surface, near the CMS or ATLAS interaction point.
We summarize the FASER geometry in table\,\ref{tab:exp_specs} (we have considered the design of FASER-2, which will have a cylindrical detector of radius of 1 m\,\cite{Ariga:2019ufm}). See instead eq.\,\eqref{eq:Mathusla_geom} below for a discussion of the configuration of the MATHUSLA detector.
In the following we consider the production of dark particles in meson decays or directly from parton collisions.\\  

\noindent\textbf{Dark particles production in mesons decays} has already been discussed in sec.\,\ref{sec:CHARM_SHIP}. We follow the same procedure also for the experiments at $\sqrt{s} = 14$ TeV. The average number of mesons produced in the collisions has been computed using {\tt EPOS-LHC} and {\tt{PYTHIA8}} and is summarized in table\,\ref{tab:multiplicities}. 
For the production of the $J/\Psi$ and $\Upsilon$ mesons, we have introduced a correction factor to the output of the {\tt{PYTHIA8}} simulations, to match the production cross-sections measured by the LHCb collaboration\,\cite{Aaij:2015rla,Aaij:2018pfp}. It is $1.6$ and $0.25$ respectively for the $J/\Psi$ and $\Upsilon$ mesons. 
A small modification of eq.\,\eqref{eq:prodMes} is needed to compute the total number of mesons produced at the experiment. The equation now reads
\be
N^{\phi_2}_{M} = {\cal L}\, \sigma_{pp}^{\rm tot} \,N_M\, {\rm BR}(M\rightarrow\phi_1\,\phi_2+{\rm X}),
\label{eq:prodMes_2} 
\ee
where ${\cal L}$ is the integrated luminosity collected by the experiment and $\sigma_{pp}^{\rm tot}$ the total proton-proton cross section at $\sqrt{s} = 14$ TeV. Using {\tt EPOS-LHC} we obtain $\sigma_{pp} \simeq 110.69$ mb, in agreement with the experimental measurement\,\cite{Antchev:2017dia} (which has been performed at $\sqrt{s} = 13$ TeV). 
The number of signal events in the detector is computed using eqs.\,\eqref{eq:total_phi2_mes},\,\eqref{eq:Ndec},\,\eqref{eq:effE} and\,\eqref{eq:prodMes_2}, and following the procedure of sec.\,\ref{sec:CHARM_SHIP}. \\

\noindent\textbf{Dark particles production from parton collisions} is computed as follows.
We implement the effective operators in eq.\,\eqref{eq:EFT} in \verb!Feynrules!\,\cite{Degrande:2014vpa} and use\\ \noindent\verb!MadGraph5_aMC@NLO!\,\cite{Alwall:2014hca} to simulate $p p \to \phi_1 \phi_2$ events at $\sqrt{s} = 14$ TeV. 
From this sample, we can compute the number of signal events in the detector using the same procedure already outlined in sec.\,\ref{sec:CHARM_SHIP}.
More in detail, eqs.\,\eqref{eq:Ndec} and \eqref{eq:effE} can be applied directly also in this case. As for the number of $\phi_2$ produced in the parton collisions, it is obtained from:
\be
N^{\phi_2}_{\rm pp} = {\cal L}\,  \sigma_{pp \to \phi_1 \phi_2},
\label{eq:prodPart} 
\ee
where the $\sigma_{pp \to \phi_1 \phi_2}$ cross section is computed with \verb!MadGraph5_aMC@NLO!. 

We must, however, ensure that the EFT in eq.\,\eqref{eq:EFT} is used inside its domain of validity.
Any EFT is a valid description of a more fundamental theory only for processes occurring at energy scales smaller than the cut-off of the EFT, $M_{\rm cut}.$ 
The latter can be written as
$M_{\rm cut}=g_*\,\Lambda,$ where $g_*$ is a combination of couplings of the UV theory, for example $g_*=\sqrt{g_{\phi}g_{V_f}}$ for the $Z^{\prime}$ model in sec.\,\ref{sec:intro} (we remind that we are taking $c_{f_R}=c_{f_L}=1$ for all the fermions).
As an example of a weakly coupled theory we take $g_*=1.$
Following\,\cite{Racco:2015dxa,Bertuzzo:2017lwt}, we then impose $\sqrt{\hat{s}} \leq \Lambda$, with $\sqrt{\hat{s}}$ the center of mass energy of the partonic event. In other words, from our sample of events simulated with \verb!MadGraph5_aMC@NLO!, we select only those satisfying this requirement.
Of course, this cut is not needed when we consider the $Z^{\prime}$ model in sec.\,\ref{sec:on_shell_med}. In that case, apart from this difference,
we implement the operators in eq.\,\eqref{eq:heavy_Zp} in \verb!Feynrules! and we compute the signal events following the same steps explained above.

Another important concern is the uncertainty in the parton distribution functions, and even the use of perturbative QCD to describe the production process.
This point can be understood as follows (see ref.\,\cite{Feng:2017uoz} for a discussion). 
For small opening angles $\theta_{\phi_2}$, (i.e. if the detector is small in the transverse direction or placed at a large distance from the interaction point) only $\phi_2$ particles with low transverse momenta reach the detector. 
This implies that one is interested in events with small momentum transfer, and the parton distribution functions has to be evaluated at low factorization scales $Q$ and momentum fractions $x$, in a regime where they suffer from large uncertainties, or even the description of the hadrons in terms of partons in perturbative QCD breaks down. 
Inspecting table\,\ref{tab:exp_specs} we see that CHARM, SHiP and FASER have very small angular acceptances, and we have checked that for these experiments only events with relatively small transverse momentum ($\lesssim 2$ GeV) are involved. 
For this reason, we do not include direct parton production in the computation of the signal events for those experiments.

The case of MATHUSLA is different: given its location, only particles with relatively large transverse momentum reach the detector on the surface.
Therefore, we consider $\phi_2$ production both via meson decays and parton collisions. 
 We show the relative importance of the various contributions in the right panel of fig.\,\ref{fig:Nmesons}. As it can be seen, for MATHUSLA parton production dominates. The geometry of the detector is the most recent one in ref.\,\cite{Alpigiani:2020tva}. The detector is delimited by:
\be\label{eq:Mathusla_geom}
68\,{\rm m} \leq z \leq 168\, {\rm m}\ , ~~~ 60\,{\rm m} \leq x \leq 80\, {\rm m}\ ,~~~ -50\,{\rm m} \leq y \leq 50\, {\rm m}\ , 
\ee
where the coordinate system is centered at the LHC interaction point, the $z$ axis is along the beam direction, and $x$ denotes the vertical to to the surface.  

We conclude mentioning that, for both FASER and MATHUSLA, the cut on the energy of the $\phi_2$ decay products discussed in sec.\,\ref{sec:CHARM_SHIP} is $E_{\rm cut} =2$ GeV, and we assume an efficiency of reconstruction of $100\%$ for all channels.

\subsection{Collider searches}\label{sec:collider_searches}
In this section we discuss constraints from accelerators which do not rely on the inelastic nature of the dark sector. More specifically we will discuss 
bounds from searches of missing energy events at LEP, LHC and BaBar, as well as limits from the invisible decays of heavy quarkonia states. 

LEP can be used to test our scenario in different ways: {\it (i)} using the excellent measurement of the $Z$ decay width to constrain exotic decay modes, and {\it (ii)} from searches of mono-photon events with large missing energy. We start by considering the first possibility. The EFT of eq.\,\eqref{eq:EFT} induces the 4-body decay $Z \to \phi_1 \phi_2 \bar{f} f$ and the invisible decay $Z \to \phi_1 \phi_2$ via a fermion loop. The corresponding decay widths scale as $\Lambda^{-4}$ and are thus suppressed at large $\Lambda$. We have used \verb!MadGraph5_aMC@NLO! to compute the 4-body process. Focusing on values of $\Lambda$ where the EFT is valid, say $\gtrsim M_Z$ (see discussion in sec.\,\ref{sec:CHARM_SHIP}),
we found that the decay width of the exotic process is always at least three orders of magnitude smaller than the current uncertainty on the $Z$ width.
Therefore, we can simply neglect this constraint.
Analogously, also the $Z$ radiative decay into $\phi_1 \phi_2$ gives a very weak bound\,\cite{Boehm:2020wbt}. 

\begin{figure}[tb]
\begin{center}
\includegraphics[width=.48\textwidth]{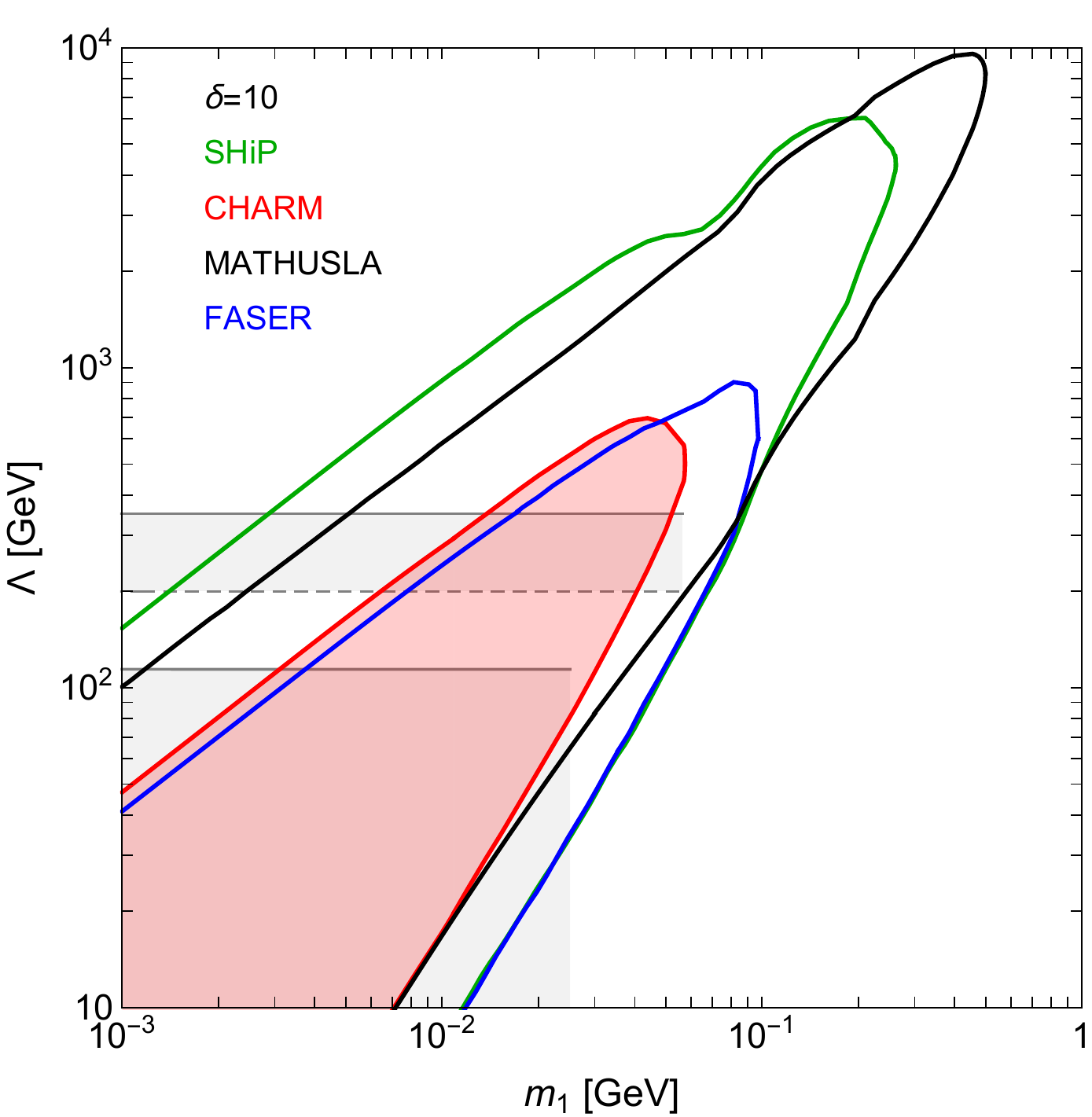}
\quad
\includegraphics[width=.48\textwidth]{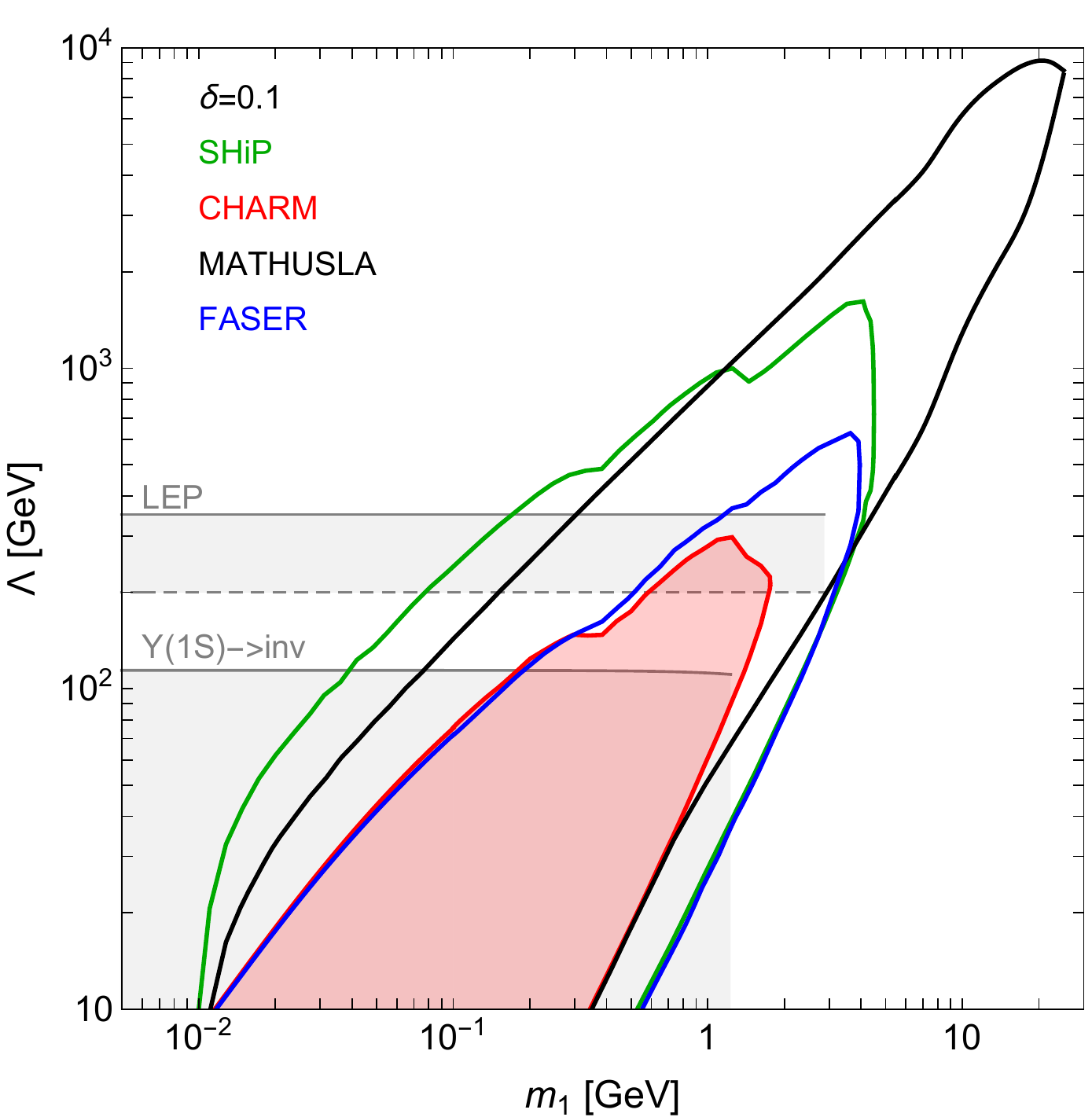} \\
\includegraphics[width=.48\textwidth]{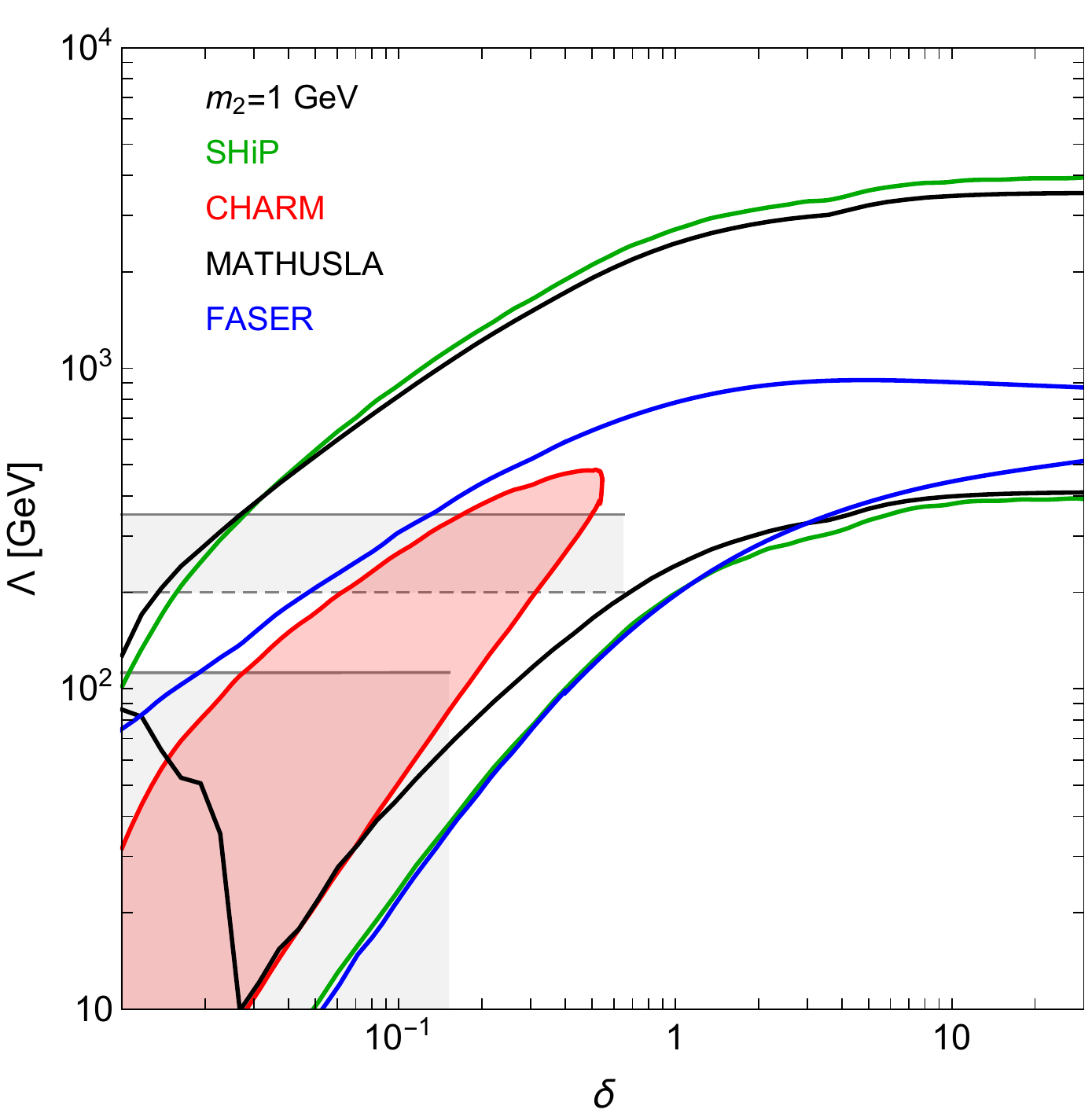}
\caption{\label{fig:Ship} \em
90\% CL exclusion limits of the CHARM experiment, and projected sensitivities of the SHiP, MATHUSLA and FASER experiments 
to the scale $\Lambda$ of EFT operators in eq.\,\eqref{eq:EFT}.
In the top panel the sensitivity reaches are shown as a function of the mass $m_1$ of the lightest dark scalar. Left (right) panel is for a mass splitting $\delta = 10$ ($\delta = 0.1$).
In the bottom panel the sensitivities are presented as a function of the mass splitting between the dark scalars, and for a mass of the heaviest one of $m_2=1$ GeV.
In all panels we take $c_{f_L} = c_{f_R} = 1$.
Gray regions are excluded by the constraints discussed in sec.\,\ref{sec:collider_searches}.
 }
\end{center}
\end{figure}

A more interesting limit is obtained from searches of mono-photon events. We consider the analysis of ref.\,\cite{Fox:2011fx}, where bounds from these searches at LEP II have been used to constrain the interaction of fermionic DM  with the SM, in the context of an EFT approach.
We recast these results for our effective operator in eq.\,\eqref{eq:EFT} with the following simplified method. In ref.\,\cite{Fox:2011fx} a bound $\Lambda > 500\,\rm{ GeV}$ has been obtained for the vector operator $ \bar{\chi}\gamma_{\mu}\chi\,\bar{e}\gamma_{\mu}e/\Lambda^2$ coupling electrons to a DM  candidate $\chi$ lighter than $(60 \div 70)$ GeV. To estimate the bound in our case, we simply factorize the production cross section as $\sigma(e^+e^-\rightarrow\bar{\chi}\chi\gamma)\simeq\sigma(e^+e^-\rightarrow\bar{\chi}\chi)\,R(e\rightarrow e\gamma)$. We take the function $R(r \to e\gamma)$ to be universal, and ignore for simplicity the difference in the angular distributions between the fermionic and scalar cases. From the scaling $\sigma(e^+e^-\rightarrow\bar{\chi}\chi)\sim1/\Lambda^4$ and the ratio of the cross sections of the processes  $e^+e^-\rightarrow\bar{\chi}\chi$ and $e^+e^-\rightarrow\phi_1\phi_2,$ we obtain the constrain $\Lambda \gtrsim350$ GeV. This is reported with a solid gray line in figs.\,\ref{fig:Mathusla} and\,\ref{fig:Ship}.
Notice however that, as already mentioned in sec.\,\ref{sec:FASER_MATHUSLA}, one should ensure that the EFT is applied inside its domain of validity.
Following the previous section and taking $M_{\rm cut}=\Lambda,$ 
we conclude that the constraint derived above for the EFT can not be consistently applied for $\Lambda$ smaller than the center of mass energy at LEP, i.e. $E_{\rm cm}\simeq200\,\rm{ GeV}.$
This lower limit is depicted with a dashed gray line in figs.\,\ref{fig:Mathusla},\,\ref{fig:Ship}. 
To test small values of $\Lambda,$ outside the validity of the EFT, it is necessary to specify the microscopic origin of the EFT. For instance, ref.\,\cite{Fox:2011fx} has explored models where the EFT operators arise from the exchange of an s-channel mediator.
We will come back to this case in sec.\,\ref{sec:on_shell_med}. Another issue that must be taken into account is that, for the mono-photon search to apply, we need to make sure that $\phi_2$ decays outside the detector. To estimate the region of the parameter space where this happens we simulate $e^+ e^- \to \gamma \phi_1 \phi_2$ events with a center-of-mass energy of $200$ GeV, and compute the average proper decay length of $\phi_2$ in the LAB frame: $c \tau_{\phi_2}\langle \gamma_{\phi_2}\beta_{\phi_2}  \rangle $. 
Requiring a proper decay length larger than the 
size the detector, which we take to be $5$ m, we obtain an upper limit $m_1 \simeq 0.06$ GeV ($2.9$ GeV) for $\delta = 10$ ($0.1$). These cuts on the mono-photon bounds are
 visible in figs.\,\ref{fig:Mathusla},\ref{fig:Ship}.

We shall now consider LHC searches. The problem of the validity of the EFT arises also in the interpretation of missing energy signatures at the LHC. Still, conservative but consistent constraints can be obtained including in the analysis of LHC data only those signal events with a center of mass energy below $M_{\rm cut}$\,\cite{Racco:2015dxa}.
This procedure has been applied in\,\cite{Bertuzzo:2017lwt} to recast mono-jet searches in the framework of the EFT of a DM singlet fermion, finding that for $g_*=1$ no bound could be set, while some region of the parameter space was probed assuming the large coupling $g_*=4\pi,$ representative of a strongly coupled UV completion.
A similar calculation could be repeated for our scenario with current LHC data. 
However, for the analysis of the LHC searches, we prefer to resort to a simple UV completion in sec.\,\ref{sec:on_shell_med}, which will allow a more pertinent investigation of current collider bounds.

Let us now move to colliders working at lower energies. We consider searches performed by BaBar in $e^+e^-$ collisions at $\sqrt{s}\simeq10\,\rm{ GeV}$, near the resonances $\Upsilon(2S),$ $\Upsilon(3S)$ and $\Upsilon(4S)$\,\cite{Lees:2017lec}. Invisible decays of heavy quarkonium states is a handle to probe light dark particles\,\cite{Fernandez:2014eja,Fernandez:2015klv}. The current upper limit on the invisible decay of the resonance $\Upsilon(1s)$ has been obtained by the BaBar collaboration, and reads $BR(\Upsilon(1s)\rightarrow{\rm inv})<3\times10^{-4}$\,\cite{Aubert:2009ae}. Using the decay width of $\Upsilon(1s)$ into $\phi_1\,\phi_2$ computed in appendix\,\ref{app:useful_eqs}, this leads to the constraint shown in figs.\,\ref{fig:Mathusla},\ref{fig:Ship}. As for the case of mono-photon searches at LEP, 
we need to ensure that $\phi_2$ decays outside the detector.
We use a procedure similar to the one described above, imposing for $\phi_2$ 
a proper decay length in the LAB frame larger than $3$ m.
The effect of this requirement can be seen in the vertical cuts in the $\Upsilon(1{\rm S}) \to {\rm inv}$ bounds in figs.\,\ref{fig:Mathusla},\ref{fig:Ship}. A less stringent bound is obtained from the upper limit on the invisible decay of the $J/\psi$\,\cite{Ablikim:2007ek}. Also searches for mono-photon events and large missing energy in $e^+e^-$ collisions could put bounds on the Wilson coefficients of the model. In our scenario, the corresponding bound turns out to be less stringent than the one derived above from the $\Upsilon(1s)$ invisible decay, in the kinematical range where the latter applies, i.e. $m_1+m_2<M_{\Upsilon}$\,\cite{Boehm:2020wbt}.

\section{Sensitivities on the EFT operators}\label{sec:bounds_EFT}

We are now in a position to discuss how the experiments described in sec.\,\ref{sec:exp} can be used to exclude or probe the parameter space of the EFT defined in eq.\,\eqref{eq:EFT}. 
Let us remind that we are taking the following choice of vector democratic couplings: $c_{f_L}=c_{f_R}=1$ for all the SM fermions.
The exclusion region from CHARM is obtained recasting the search in ref.\,\cite{Bergsma:1985is} where no events have been observed. A limit at 90\% CL can be obtained imposing that the number of signal events is $N_{\rm sig}<2.3.$
The same criterion is used to derive the sensitivity reach of the SHiP, FASER and MATHUSLA experiments. This assumes that backgrounds can be reduced at a negligible level, which is expected to be the case according to present studies\,\cite{SHiP:2018yqc,Alpigiani:2020tva,Ariga:2019ufm}.
Moreover, our sensitivity contours are only marginally affected by an imperfect reconstruction of the signal (i.e. assuming an efficiency $\epsilon_i<1$), or a moderate number of background events.
This is due to the steep dependence of the number of signal events on $\Lambda.$ In fact, the number of $\phi_2$ particles produced in meson decays or via direct parton collisions scales as $\Lambda^{-4}.$ Moreover, the fraction of those events which decay inside the detector depends exponentially on the $\phi_2$ lifetime, see eq.\,\eqref{eq:Ndec}, which in turn grows like $\Lambda^{4.}$

Let us now present our results in the $(m_1, \Lambda)$ plane for the two representative cases of a large, $\delta=10,$ and small, $\delta=0.1,$ mass splitting among the dark states. First we show our sensitivity contours for MATHUSLA in fig.\,\ref{fig:Mathusla}, including separately only events from meson decays (green line), direct parton production (red line), and then combining the two processes (black line).
As expected from fig.\,\ref{fig:Nmesons} (right panel), parton production is generically more relevant. In particular it allows to extend the reach at masses of the dark states that can not be explored with meson production for kinematical reasons (simply because $m_1+m_2$ is larger than the mass of the mesons).
However, there are regions of the parameter space where the two processes give comparable sensitivities, or the production via meson decays is more relevant.
The reason for that is the term in eq.\,\eqref{eq:Ndec}, which accounts for the probability that the $\phi_2$ states decay inside the detector.
As discussed below eq.\,\eqref{eq:Ndec}, it depends on the distance travelled by the $\phi_2$ particle, and therefore its Lorentz factor $\gamma_{\phi_2}$. Production from meson decays or via direct parton collisions lead to different distributions for $\gamma_{\phi_2}$, centred around larger values for the latter process.
Therefore the term in eq.\,\eqref{eq:Ndec} can be very different in the two cases.
Notice also, from the right panel of fig.\,\ref{fig:Nmesons}, that the number of dark pairs produced by heavy meson decays is larger than those from the lighter ones.

One can notice in fig.\,\ref{fig:Mathusla} that for a given mass $m_1,$ MATHUSLA is able to probe a limited range of $\Lambda.$ For small values of $\Lambda$ the $\phi_2$ particles decay before reaching the detector, while in the opposite case their production is suppressed, and/or they decay at large distances.
Moreover, for the direct parton production, small values of $\Lambda$ are not included inside our sensitivity reach, see the right panel of fig.\,\ref{fig:Mathusla}.
This is because the cut introduced in sec.\,\ref{sec:FASER_MATHUSLA} to ensure the validity of the EFT approach, removes most of the signal events in this region.

In fig.\,\ref{fig:Ship} we compare the sensitivities of the different experiments that we have analyzed.
The constraints discussed in sec.\,\ref{sec:collider_searches} are shown in gray. The left panel is for $\delta = 10.$ As evident, the CHARM exclusion region extends well above the bounds from LEP. FASER will be able to further improve these limits. 
SHiP and MATHUSLA can probe a much larger region of parameter space, reaching up to $\Lambda \simeq (5 \div 8)$ TeV for $m_2 \sim (1 \div 3)$ GeV. 
In the case of SHiP (and less pronounced for FASER) a feature is visible around $m_2 \simeq 0.6$ GeV, which corresponds to the threshold at which the main production channel for $\phi_2$ moves from $\omega$ decays to J/$\Psi$ decays. 

We now turn to the right panel, with $\delta = 0.1$. Qualitatively, the results are similar to the previous case, but smaller values of $\Lambda$ can be probed. 
The effect is particularly evident in the case of SHiP and, to a lesser extent, for CHARM.
Consider a fixed value of $\Lambda,$ for instance $\Lambda=2$ TeV. With this small mass splitting, $\phi_2$ has a large decay length, which causes most of the decays to happen after the detector, and diminishes the number of events. The decay length can be reduced increasing the mass of the scalars, but if they are too heavy they can not be produced in meson decays.
Again, the features appearing in the contour regions of SHiP, FASER and CHARM, signal the transitions from $\omega,$ to J/$\Psi$ and $\Upsilon$ decays as the main production mechanism for $\phi_2.$
A difference with respect to the left panel is the range of $\phi_1$ masses which can be probed.
In this case, with a quite compressed mass spectrum, larger dark scalar masses are needed to lead a detectable signal through the process $\phi_2 \to \phi_1 + {\rm X_i}.$
For $m_1\lesssim10^{-2}$ GeV even the electron channel (${\rm X_i}=e^+e^-$) is closed, and $\phi_2$ can only decay into neutrinos. To give an idea of the lifetimes that can probed with these experiments, we compute $c \tau_{\phi_2}$ in the rest frame of the $\phi_2$ particle for several choices of the dark particles masses and $\Lambda$. 
For $\delta=10$, we obtain $c\tau_{\phi_2} = 5.6\times 10^4\,{\rm m}$ for  $(m_1, \Lambda) = (2\times 10^{-3} \,{\rm GeV}, 100\,{\rm GeV})$ and $c\tau_{\phi_2} = 1.4\,{\rm m}$ for $(m_1, \Lambda) = (0.3 \,{\rm GeV}, 6\,{\rm TeV})$. Notice that it is trivial to rescale these results for other values of $\Lambda$, since the decay length simply scales as $c\tau_{\phi_2} \propto \Lambda^4$. Turning to $\delta= 0.1$, we find $c\tau_{\phi_2} = 2.5 \times 10^5\, {\rm m}$ for $(m_1, \Lambda) = (0.1\,{\rm GeV}, 100\,{\rm GeV})$ and $c\tau_{\phi_2} = 4.2\, {\rm m}$ for  $(m_1, \Lambda) = (10\,{\rm GeV}, 3\,{\rm TeV})$. As evident, future experiments will be able to probe decay lengths that span many orders of magnitude. 
We remind that in the computation of the experimental sensitivities one should property take into account the relativistic $\gamma_{\phi_2}$ factor to determine the decay length of $\phi_2$ in the LAB frame.
This has been done as explained in sec.\,\ref{sec:CHARM_SHIP}.

In the bottom panel of fig.\,\ref{fig:Ship}, we the show a different slice of the parameter space. We fix the mass of the heaviest dark scalar $M_2=1$ GeV, and we present our sensitivities as a function of the mass splitting $\delta.$ The contours saturate at large values of $\delta$ simply because $\phi_1$ can be considered effectively massless for large enough $\delta.$
Notice also that there is a region of the parameter space at small $\delta$ and $\Lambda$ which is not covered by MATHUSLA.
In this corner of the parameter space, the production of dark states via meson decays do not lead to detectable signals because the $\phi_2$ decay products are too soft to satisfy our cut on their energy in sec.\,\ref{sec:FASER_MATHUSLA}. The situation is different for FASER, CHARM and SHiP, since these detectors are placed along the beam axis, and they are 
typically crossed by $\phi_2$ particles with larger energy, for which it is easier to produce decay products that satisfy the energy cut.
For events produced directly from parton collisions, the cut on the center of mass energy of the partonic event, $\sqrt{\hat{s}}<\Lambda$ (see sec.\,\ref{sec:FASER_MATHUSLA}), becomes increasingly more stringent at small $\Lambda$, and selects events with softer $\phi_2$'s. Then, these low energy events are typically removed by the requirement on the energy of the $\phi_2$ decay products.

Let us close this section mentioning some other relevant experimental results.
New limits on dark particles have been obtained by the MiniBooNE collaboration from a search performed with an $8$ GeV proton beam dump\,\cite{Aguilar-Arevalo:2018wea}. We have recasted these results for our scenario with a simplified method to implement the experimental analysis cuts.
We have found that the region of the parameter space excluded by this search (for the same benchmark scenarios in fig.\,\ref{fig:Ship}) is already tested by the CHARM experiment. A more detailed analysis would be necessary to outline precisely the exclusion limits.
Other complementary constraints could be  derived from searches at LSND\,\cite{Auerbach:2001wg}.
Searches of missing energy in the electron fixed target experiment NA64\,\cite{NA64:2019imj} leads to weak bounds in our scenario, see\,\cite{Boehm:2020wbt}.
We have also estimated the constraints from the electron beam dump E137\,\cite{Bjorken:1988as}. We have found that it probes a region of parameter space already excluded by CHARM and  the collider bounds  in fig.\,\ref{fig:Ship}.

\section{Sensitivities on the $Z'$ model}\label{sec:on_shell_med}

We now focus on a simplified model that provides a partial UV completion for the EFT defined in eq.\,\eqref{eq:EFT}.
Specifically, we are considering the $Z^{\prime}$ model introduced in sec.\,\ref{sec:intro}.
Our aim is to properly study the sensitivities of the various experiments under consideration, in the case where the mediator of the interaction among the dark states and the SM can be produced on-shell. In this situation, as mentioned in sec.\,\ref{sec:collider_searches}, the use of a simplified model allows us an in-depth comparison with the bounds from high-energy accelerators, in particular with those from the ATLAS and CMS experiments at the LHC. 
We consider two benchmark cases: a heavy $Z'$ (with a mass above 1 TeV) coupled to the SM fermions via vector currents, and a light dark photon (with a mass of $40$ GeV) interacting with the SM via kinetic mixing (see for instance\,\cite{Curtin:2014cca}). 

\subsection{A heavy $Z'$}
\label{sec:heavyZp}
\begin{figure}[tb]
\begin{center}
\includegraphics[width=.48\textwidth]{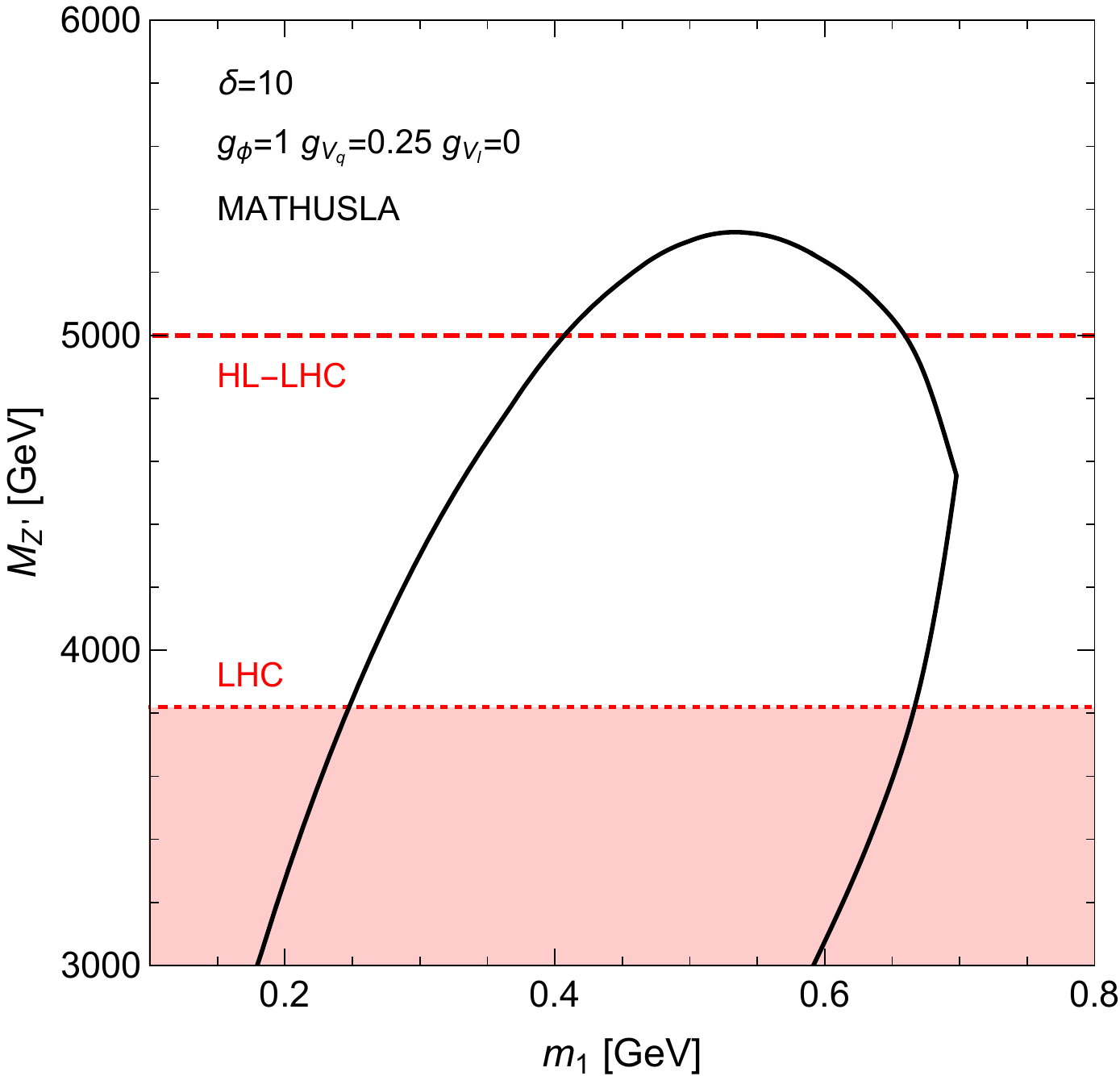}
\quad
\includegraphics[width=.48\textwidth]{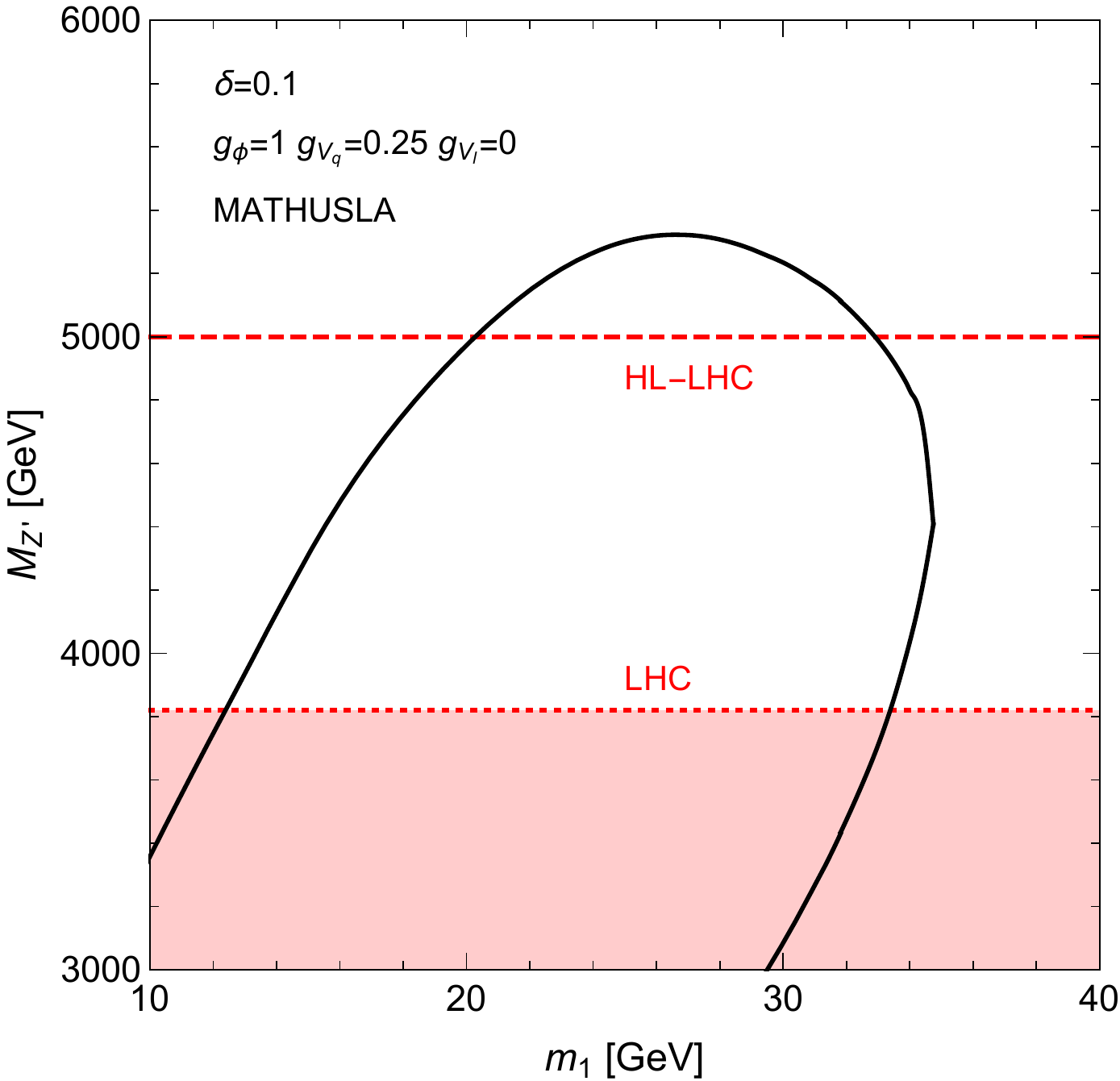} \\
\includegraphics[width=.48\textwidth]{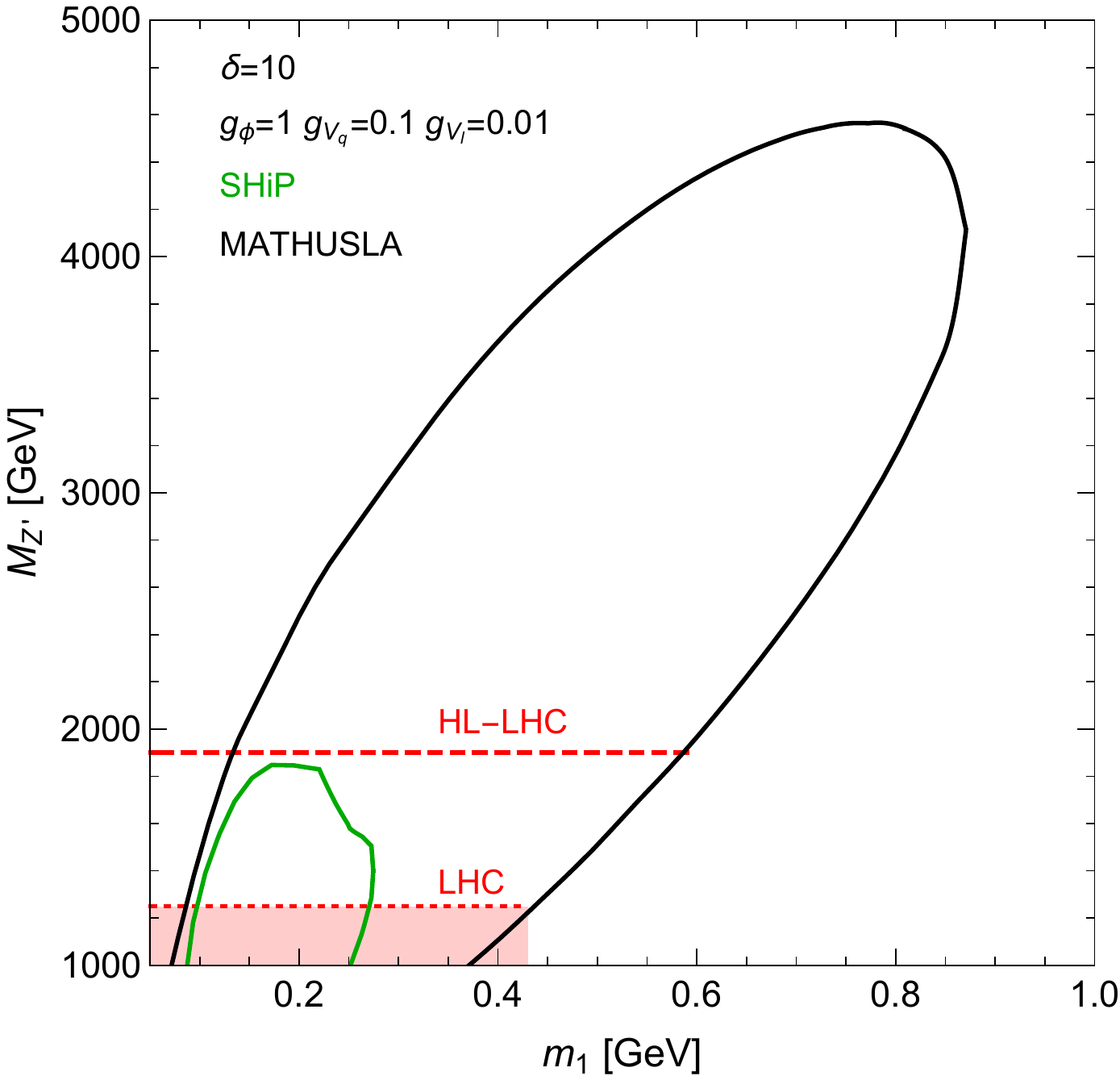} 
\quad
\includegraphics[width=.48\textwidth]{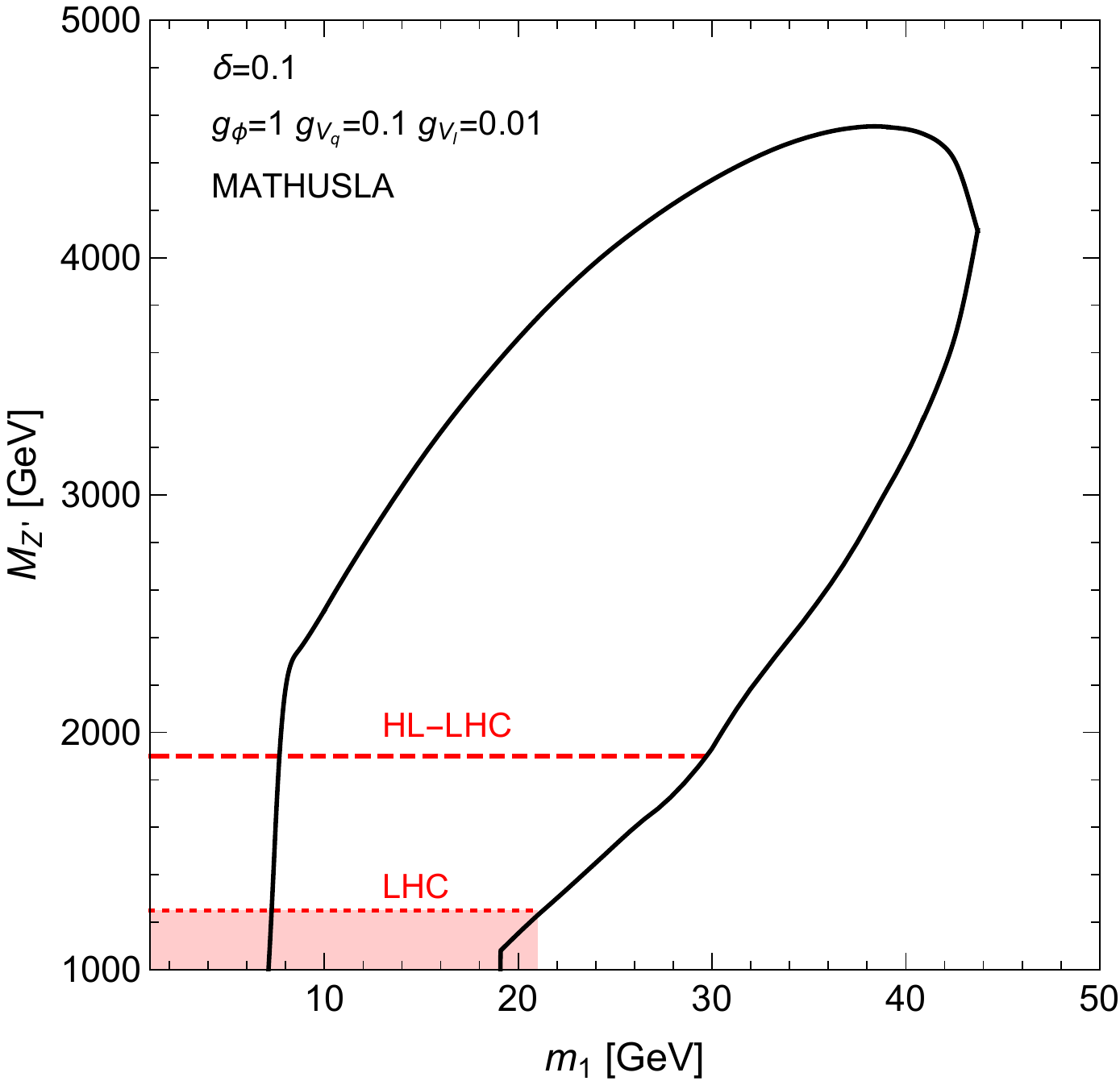}
\caption{\em \label{fig:Zp_heavy} 
90\% CL projected sensitivities to the $Z^{\prime}$ model in eq.\,\eqref{eq:heavy_Zp}.
Top and bottom panels are for two choices of the couplings of the $Z^{\prime}$ to the SM fermions.
Left and right panels are for a mass splitting between dark states of $\delta=10$ and $\delta=0.1$ respectively.
The red regions below the red dotted line are excluded by the current LHC constraints discussed in sec.\,\ref{sec:heavyZp}.
The dashed red lines show the expected future sensitivity of the HL-LHC.
 }
\end{center}
\end{figure}
The simplified model that we are considering is specified by the lagrangian in eq.\,\eqref{eq:heavy_Zp}. 
At low energies, these interactions lead to the EFT in eq.\,\eqref{eq:EFT}, with the Wilson coefficients:
\be
\label{eq:maching}
\frac{c_{f_L}}{\Lambda^2} = \frac{g_{\phi}\, g_L^f}{M_{Z^\prime}^2}\ , ~~~ \frac{c_{f_R}}{\Lambda^2} = \frac{g_{\phi} \,g_R^f}{M_{Z^\prime}^2}\ ,
\ee 
where $M_{Z^\prime}$ is the mass of the $Z^{\prime}$ boson.
Since in this section we are considering  $M_{Z^\prime} \gtrsim 1$ TeV, for the CHARM, SHiP and FASER experiments we can apply the analysis on the EFT operators explained in sec.\,\ref{sec:bounds_EFT} and use eq.\,\eqref{eq:maching}~\footnote{In the case of the FASER experiment the use of the EFT is justified since we consider only $\phi_2$ production via meson decays, see sec.\,\ref{sec:FASER_MATHUSLA}.}.

For the MATHUSLA experiment the situation is different, since the $Z^{\prime}$ boson can be produced on shell at the LHC, a situation which is not captured by the EFT.
Therefore, we implement the interactions of eq.\,\eqref{eq:heavy_Zp} in \verb!Feynrules!\,\cite{Degrande:2014vpa} and use \verb!MadGraph5_aMC@NLO!\,\cite{Alwall:2014hca} to simulate $p p \to Z' \to \phi_1 \phi_2$ events at $\sqrt{s} = 14$ TeV.
The analysis then proceeds as explained in sec.\,\ref{sec:FASER_MATHUSLA}.

We focus on vectorial couplings, defined in eq.\,\eqref{eq:vector_axial_coupls_Zp}, and we consider two benchmark scenarios. In the first case, the $Z^{\prime}$ is hadrophilic and its couplings to the SM quarks and leptons are respectively $g_{V_q} = 0.25$, $g_{V_l} = 0.$
In the second scenario an interaction with the leptons is turned on: $g_{V_q} = 0.1$, $g_{V_l} = 0.01$. In both cases the interaction is flavor blind (same coupling for all the flavors), and $g_{\phi}=1.$
These choices are motivated by the possibility to confront with the LHC searches presented in ref.\,\cite{ATLAS:2020fmm}.
For the hadrophilic scenario, the strongest constraint is from searches of a new resonance using dijets.
In the other case, the best limit is obtained through searches of  missing energy and photons or jets ($E_T^{\rm miss} + X$ in \,\cite{ATLAS:2020fmm}).
For these analysis, and for the range of masses of the dark scalars that we are considering, the dark states can be effectively considered massless, and the mass splitting plays no role.
The constraints in\,\cite{ATLAS:2020fmm} are presented for a simplified model including a Dirac dark fermion, besides the $Z^{\prime}$ boson. Instead, we are considering  
dark scalars. We can recast these bounds for our scenario using the cross-sections of the processes $p\,p\rightarrow Z^{\prime}\rightarrow j\,j,$ $p\,p\rightarrow Z^{\prime}\rightarrow j\,\phi_1\phi_2$, and the analogous ones for the Dirac dark fermion\footnote{Note that the nature of the dark particles, fermion vs scalar, is also relevant for the search of dijets, since it affects the total width of the $Z^{\prime}$ boson.}.
With this approximate procedure, we find the exclusion limits presented as dotted lines in fig.\,\ref{fig:Zp_heavy}. 
They correspond to a shift of $\simeq5\%$ and $\simeq9\%$ with respect to the bounds for the Dirac fermion, respectively for the first and second scenarios.
Following a similar method, and assuming that future constraints on the signal cross-sections will improve as the square root of the luminosity, we estimate the reach of future searches at the High-Luminosity LHC (HL-LHC), with an integrated luminosity of $3$ ab$^{-1}.$
The results are shown in fig.\,\ref{fig:Zp_heavy} with dashed lines. Care must be taken in the interpretation of the bounds  from $E_T^{\rm miss} + X$. The problem is the same already discussed in sec.\,\ref{sec:collider_searches} for LEP and Babar: one should ensure that the $\phi_2$ lifetime is long enough such that it decays outside the detector. 
To estimate the region of the parameter space where this happens we follow the same procedure outlined for LEP, requiring the proper decay length of $\phi_2$ in the LAB frame to be larger than the detector radius, which we take equal to $10$ m. This explains the vertical cuts in the excluded regions 
in the lower panels of fig.\,\ref{fig:Zp_heavy}. Other searches (like prompt or displaced decays) may be relevant in the region of the parameter space where $\phi_2$ decays inside the detector. Although interesting, they lie beyond the scope of the paper.

Now we compare these searches at the LHC with the sensitivities of the CHARM, SHiP, FASER and MATHUSLA experiments.
Our 90\% CL sensitivity limits are presented in fig.\,\ref{fig:Zp_heavy}. The top (bottom) panel refers to the first (second) benchmark scenario, and the left (right) panel is for $\delta=10$ ($\delta=0.1$).
We find that the CHARM and FASER experiments are able to test $Z^{\prime}$ masses smaller those shown in the plot, and therefore already excluded by the current bounds from LHC. 
The same is true for SHiP, except for the case in the bottom right panel of fig.\,\ref{fig:Zp_heavy}. For this choice of the parameters, SHiP will be able to improve present constraints and will be competitive with future HL-LHC limits. For the same couplings, but a smaller mass splitting (bottom right panel of fig.\,\ref{fig:Zp_heavy}), $\phi_2$ only decays into SM leptons and $\phi_1$, since hadronic decays only occurs for masses such that the dark scalars can not be produced in meson decays.
These leptonic decays are suppressed for our choice of couplings ($g_{V_q} = 0.1$, $g_{V_l} = 0.01$), and it turns out that the SHiP sensitivity falls below the current LHC constraints.

The situation is drastically different for MATHUSLA. For all the four cases in fig.\,\ref{fig:Zp_heavy}, MATHUSLA will be able to surpass current LHC limits, and even probe regions of the parameter space outside the reach of the HL-LHC. This is especially the case for the second benchmark scenario, where the couplings of the $Z^{\prime}$ boson to quarks are reduced.
Already in fig.\,\ref{fig:Ship} for the case of the EFT operators, one can notice that MATHUSLA is able to significantly extend the reach of the other experiments under consideration at large $\Lambda.$ 
The examples studied here show that MATHUSLA is very useful to explore dark sectors, and they underline its complementary to other LHC experiments, namely CMS and ATLAS. 

Let us conclude mentioning the constraints from electroweak precision measurements (EWPM). Using the results in ref.\,\cite{Cacciapaglia:2006pk} one can verify that in our  
scenario only the following electroweak oblique parameters are generated: $W$, $Y$, $X$ and $V$. The future reach of the HL-LHC on the $W$ and $Y$ parameters have been computed in\,\cite{Ricci:2020xre} analyzing the reach on Drell-Yan processes. For a coupling $g_{V_l} = 0.01$ we obtain that the region that can be probed is $M_{Z^\prime} \lesssim 325$ GeV, not competitive with current and future direct searches.

\subsection{A light dark photon}
\label{sec:DP}
\begin{figure}[tb]
\begin{center}
\includegraphics[width=0.48\textwidth]{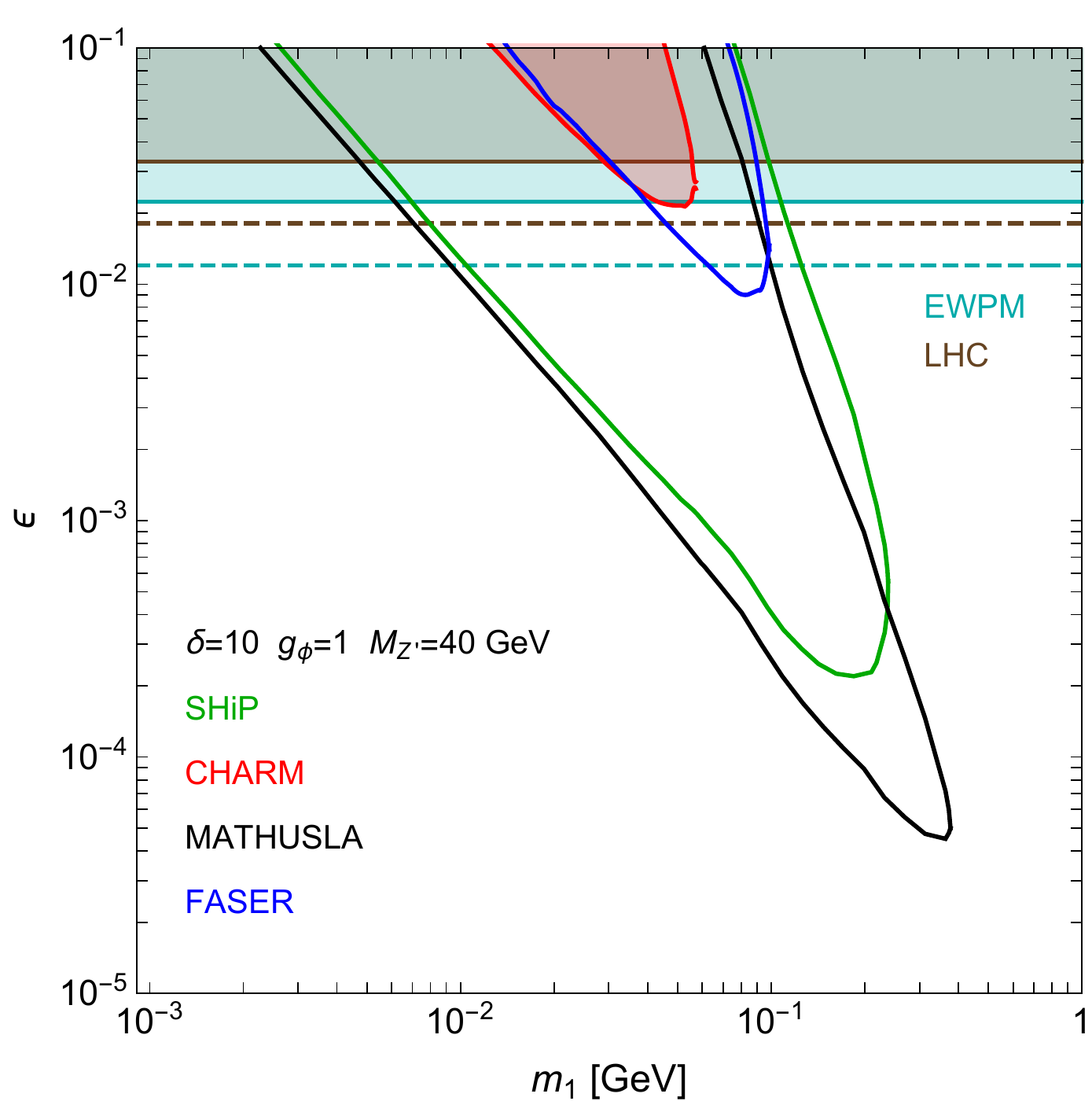}
\quad
\includegraphics[width=0.48\textwidth]{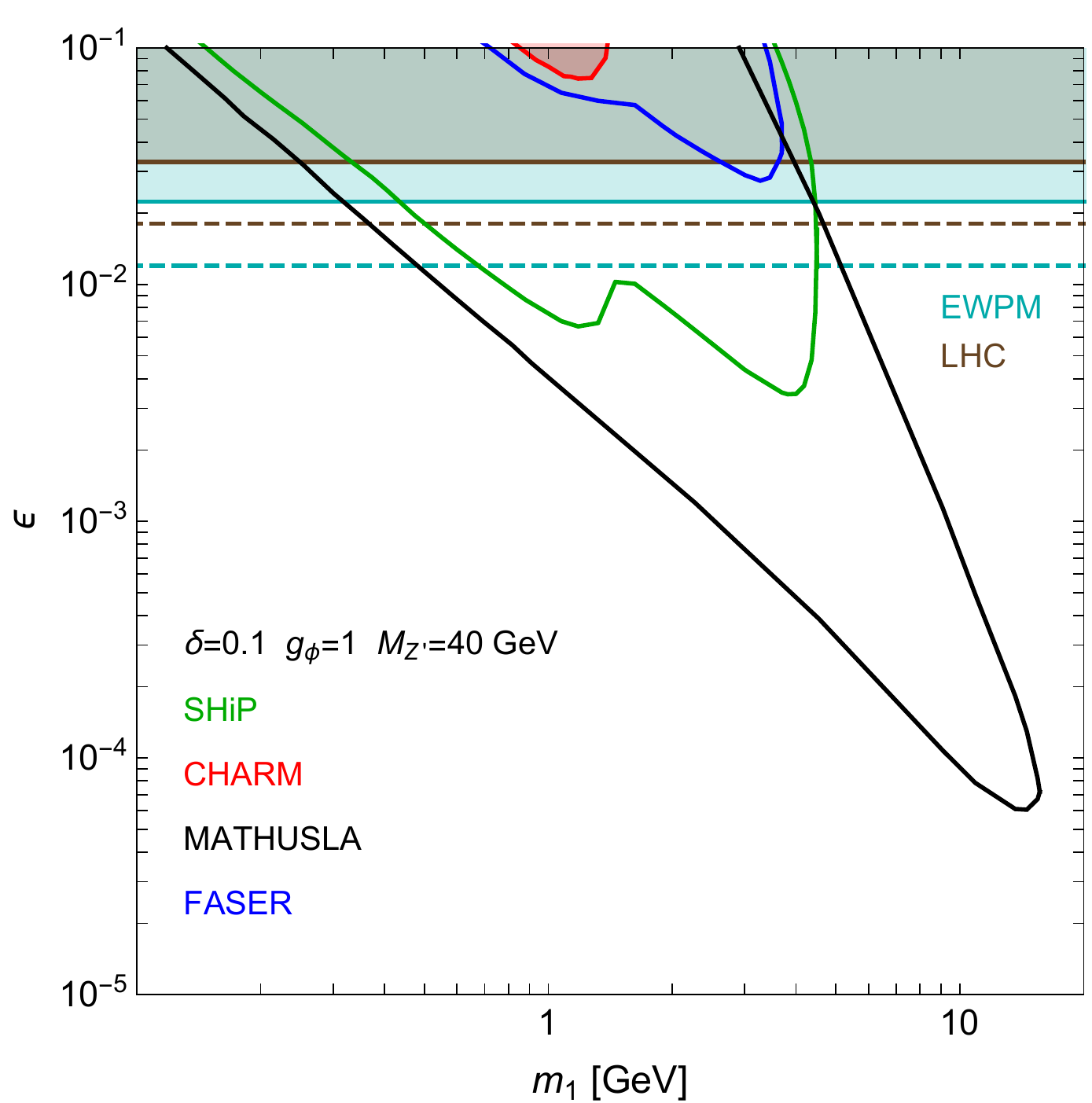}
\caption{\label{fig:dark_photon} \em
90\% CL projected sensitivities (for SHiP, MATHUSLA and FASER) and exclusion limits (for CHARM ) for the scenario with a dark photon mediator discussed in sec.\,\ref{sec:DP}.
The mass of the dark photon is taken $M_{Z^{\prime}}=40$ GeV, and the sensitivity reaches on the kinetic mixing parameter $\epsilon$ are shown as a function of the mass $m_1$ of the lightest dark scalar.
Left (right) panel is for a mass splitting between the dark scalars of $\delta=10$ ($\delta=0.1$).
Cyan and gray regions are excluded by EWPM and LHC searches discussed in sec.\,\ref{sec:DP}.
Dotted cyan and gray lines show the expected improvement of such searches.
}
\end{center}
\end{figure}
We now analyze the case of a relatively light vector boson. Specifically we focus on a dark photon, which has been extensively been studied as a portal through
which the dark and the visible sector communicate.
This particle interacts with the SM via the kinetic mixing operator\,\cite{Holdom:1985ag,Galison:1983pa,Foot:1991kb}:
\be
\mathcal{L} \supset \frac{\epsilon}{2\,c_w} F^{\prime}_{\mu\nu} B_{\mu\nu}\ ,
\ee 
where $B_{\mu\nu}$ and $F^{\prime}_{\mu\nu}$ are the field strengths of the hypercharge and the new vector boson respectively.
The adimensional parameter $\epsilon$ controls the kinetic mixing, and $c_w$ is the cosine of the Weinberg angle.
After a suitable field redefinition which diagonalize the kinetic and mass terms of the vector bosons, the dark photon acquires a coupling with the SM fermions. At leading order in 
$\epsilon\ll1$ and in $M_{Z^{\prime}}/M_Z$ ($M_Z$ is the mass of the SM $Z$ boson) these couplings are\,\cite{Curtin:2014cca}:
\be
g_{V_f} = \epsilon \,e \, Q_f\ , ~~~~ g_{A_f} = 0 \,
\ee
where $e\,Q_f$ the electric charge of fermion $f$.

As mentioned before, the dark photon has been widely studied in the literature, and relevant constraints for this scenario have been derived. We now discuss the most stringent ones focusing on the benchmark case of a dark photon mass of $M_{Z^\prime} = 40 $ GeV. A bound from EWPM has been computed in ref.\,\cite{Curtin:2014cca}. For $M_{Z^\prime} = 40 $ GeV it reads $\epsilon \lesssim 2.2\times 10^{-2}$. 
Additional limits come from searches for a light resonance decaying into leptons at LHCb\,\cite{Aaij:2019bvg} and CMS\,\cite{Sirunyan:2019wqq}.
They lead to a similar bound: $\epsilon^2 \lesssim 3 \times 10^{-6}$ for $M_{Z^\prime} = 40 $ GeV.
However, these analysis can not be applied directly to our case, since they assume that the dark photon decays into SM states only. 
We should instead take into account its decays into dark sector particles. 
This can be done computing the branching fraction for decays into leptons for the two cases where the dark photon couples to the dark sector or only to SM particles. The bound can be rescaled using the ratio of these quantities. Explicit formulas are presented in appendix\,\ref{app:Zp_decay}. 
More specifically we write:
\be\label{eq:rescaling_DP}
\epsilon^2 \frac{{\rm BR}(Z^\prime \to  l^+ l^-)}{{\rm BR}(Z^\prime \to  l^+ l^-)|_{g_{\phi} = 0}} \lesssim 3\times 10^{-6} 
\ee
and for $g_{\phi}=1$ we obtain $\epsilon \lesssim 3.3 \times 10^{-2}$, comparable to the bound obtained from EWPM. 
Constraints from searches of of mono-photon events with large missing energy at LEP have been derived in \,\cite{Ilten:2018crw}, using the analysis of \,\cite{Fox:2011fx}.
For our case the limit is comparable but less stringent than the previous ones.
Future EWPM are expected to improve the sensitivity on $\epsilon$ by about a factor of $2$\,\cite{Curtin:2014cca}. 
We use ref.\,\cite{CidVidal:2018eel} to derive future bounds from LHCb (see their figure 3.4.1).
All these constraints and sensitivities are shown in fig.\,\ref{fig:dark_photon}. There we also present the sensitivities for CHARM, SHiP,  FASER and MATHUSLA, for the two representative values of the mass splitting among the dark scalars adopted in the previous sections: $\delta=10$
(left panel) and $\delta=0.1$ (right panel).
These contours are derived with the same procedure outlined in sec.\,\ref{sec:heavyZp}.
For $\delta=10$, the current exclusion region from CHARM is comparable to the limit from EWPM, $\epsilon \sim (2\div 3) \times 10^{-2}$, but for a rather limited range of $\phi_1$ masses around $m_1 \sim 0.05$ GeV.
The reach of the other experiments extends well below the current excluded region, probing kinetic mixing down to $\epsilon \simeq 5\times 10^{-5}$ in the case of MATHUSLA.
For $\delta=0.1,$ smaller values of $\epsilon$ can be tested, analogously to the situation presented for the EFT in sec.\,\ref{sec:bounds_EFT}. Still large regions of the parameter space can be probed with SHiP and MATHUSLA.
Once again, these results demonstrate that these future experiments are useful to search for the dark sector under consideration. A similar complementarity of bounds has been explored in the case of inelastic fermion dark matter interacting with a dark photon in\,\cite{Izaguirre:2017bqb,Berlin:2018jbm,Duerr:2019dmv}.

We conclude this section pointing out that 
the dark scalar responsible for the mass of the dark photon is expected to have a mass around $M_{Z^\prime}$. For simplicity we  assume that this state is sufficiently decoupled and/or weakly coupled to the dark and visible sector to be ignored.

\section{Bounds from astrophysics and cosmology}
\label{sec:cosmo}

In this section we present astrophysical and cosmological bounds that can be relevant for the scenario considered in this paper.

\subsubsection*{DM abundance}
Although not essential, we are assuming throughout our work that $\phi_1$ is a stable particle. This is an attractive possibility since $\phi_1$ could then play the role of DM.
Let us briefly discuss few scenarios which could determine its cosmological abundance.
Assuming that DM annihilations in the early Universe are dominated by the processes induced by the effective operator in eq.\,\eqref{eq:EFT}, one can compute the DM relic density after its thermal freeze-out.
It turns out that, under this assumption, DM is overabundant in most of the parameter space that we are studying\,\cite{Boehm:2020wbt,Choudhury:2019tss}. This is because, for large enough $\Lambda$, the annihilation cross-section is too small to efficiently deplete the dark matter density.

On the other hand, the dark sector might be richer of what we are assuming, and include other states into which DM can annihilate. Still, even decoupled (or very weakly coupled) dark sectors are subject to cosmological constraints. For instance, the abundance of the additional dark states should not excessively alter the expansion of the Universe during the Big Bang Nucleosynthesis (BBN)\,\cite{Scherrer:1987rr,Hufnagel:2017dgo}. Moreover, cosmic microwave background (CMB) observations strongly constrain extra energy deposition into the primordial plasma due to annihilations or decays of dark states into SM particles, e.g.\,\cite{{Adams:1998nr,Chen:2003gz,Padmanabhan:2005es,Slatyer:2009yq,Galli:2009zc,Finkbeiner:2011dx}}.
Several possibilities exist to satisfy the cosmological bounds. Examples of those are models where DM annihilates into a stable dark sector particle (which has a small cosmological abundance)\,\cite{Duerr:2018mbd}, or where the processes determining the DM relic abundance involve dark states almost degenerate in mass with the DM, and they are suppressed at later times\,\cite{DAgnolo:2015ujb,DAgnolo:2017dbv,Dror:2016rxc,Kopp:2016yji}.

Alternatively, scenarios with an excessive DM abundance might be brought in agreement with observations if a period of entropy dilution arises after the DM freeze-out, diluting the DM abundance, see e.g.\,\cite{Evans:2019jcs} for light DM interacting through an heavy mediator.

Finally, let us remind that the predictions for the DM the relic abundance can be dramatically altered if the thermal history of the Universe before BBN was different from what is usually assumed. 
For instance, this happens if the reheating temperature of the Universe was small enough to prevent the DM thermalization with the SM plasma\,\cite{Giudice:2000ex,Chu:2013jja}. In this case, the abundance is set by the freeze-in mechanism rather than the freeze-out.

Since it is not fundamental for our discussion, in this work we do not focus on a specific model where the correct DM abundance can be obtained.
It would be interesting to explore such possibility in a future work.

\subsubsection*{BBN and CMB constraints}
DM with $\mathcal{O}({\rm MeV})$ masses in thermal equilibrium with the SM plasma at the epoch of the neutrino decoupling ($T_D\approx2.3$ MeV) 
are constrained by BBN and CMB observations. The reason is that these particles modify the expansion of the Universe and change the ratio between neutrino and photon temperatures, altering the production of light elements during BBN, and shifting the value of the effective number of neutrinos ($N_{\rm eff}$), which is inferred quite precisely from CMB measurements.
Current bounds for complex scalar DM exclude masses below few MeV, the precise value depending on the cosmological dataset considered, and the relative size of the DM annihilation cross-section into neutrinos and electrons or photons\,\cite{Boehm:2013jpa,Escudero:2018mvt,Sabti:2019mhn,Depta:2019lbe}. 
For $\Lambda\gtrsim$ few $\mathcal(100)$ GeV and a negligible mass splitting $\delta$, the dark scalars are decoupled from the SM bath at $T_D.$ Therefore BBN bounds do not apply in these regions of the parameter space.
Notice also that the lifetime of $\phi_2$ is shorter than one second in the regions of the parameter space probed by the CHARM, SHiP, MATHUSLA and FASER experiments, and shown in figs.\,\ref{fig:Ship},\,\ref{fig:Zp_heavy} and\,\ref{fig:dark_photon}. This implies that $\phi_2$ decays before the BBN epoch avoiding, for example, potential bounds related to the photodissociation of light elements from its decays.

As discussed before, CMB measurements also constrain DM annihilations into SM particles at early times. In our scenario $\phi_2$ decays before the recombination era, so the relevant annihilation processes involves only pair annihilations of $\phi_1$ into two (induced at loop level) or four SM particles.
The bounds on these processes are very weak. It is also worth mentioning that $\phi_1\phi_2$ (co)-annihilations are p-wave suppressed at low velocities.
For the same reasons, indirect dark matter searches give weak limits.

As mentioned before, additional ingredients should be added to our framework in order to obtain the correct DM abundance. Depending on the concrete model, the constraints discussed in this section might apply. \\ 

\subsubsection*{Supernova bounds}
Observations of supernova explosions are a precious tool to test light dark sectors weakly coupled to the SM. In the dense and hot supernova environment, with temperatures of the order of few tens of MeV, dark particles with masses up to $\lesssim\mathcal{O}(0.1)$ GeV can be produced.
If these particles interact weakly enough with the supernova material, they can efficiently escape, providing therefore an additional mechanism to cool the supernova. On the other hand, if the dark particles interacts too much, they are trapped inside the core of the supernova, and the cooling rate becomes negligible with respect to the one due to the neutrino emission.
Therefore, there is an optimal range of couplings between the dark particles and the SM that can be tested with this argument, for which the dark states are abundantly produced, and they efficiently escape from the supernova.
The observed cooling time of the supernova 1987A has been used to set bounds on the pair production of light DM\,\cite{Dreiner:2003wh,Fayet:2006sa,Dreiner:2013mua,Zhang:2014wra,Guha:2015kka,Tu:2017dhl,Mahoney:2017jqk,Knapen:2017xzo,Chang:2018rso,Guha:2018mli,DeRocco:2019jti,Boehm:2020wbt}.
Ref.\cite{Boehm:2020wbt} considered a complex scalar coupled to electrons through an heavy $Z^{\prime}$, leading to the effective vector operator in eq.\,\eqref{eq:EFT}.
They found that supernova limits extend up to masses of $\simeq$ 0.2 GeV, testing $\Lambda\simeq1\rm{TeV}$ at this mass scale, where the excluded region closes.
Our model differs from the one in\,\cite{Boehm:2020wbt} in several respects. We are focusing on a democratic coupling of the dark scalars to all the SM fermions. This implies that additional processes should be consider for the production of the dark particles and their interaction with the medium\,\footnote{Notice however that, in the context of a dark fermion interacting with the SM through a dark photon mediatior, ref.\,\cite{DeRocco:2019jti} found that electron-positron annihilations dominates over the production through nucleon-nucleon bremsstrahlung.}. 
Moreover, in our scenario the dark scalars present a mass splitting. This is crucial to determine whether these particles are trapped or not inside the supernova. Once produced, $\phi_2$ decays into $\phi_1$ and SM particles, and if the process is fast enough, only a population of $\phi_1$ remains.
The lightest scalar $\phi_1$ can interact with the ambient particles scattering into $\phi_2.$ However, if the mass splitting is large, this process is strongly suppressed.
Therefore the bounds at large couplings (i.e. small $\Lambda$) are completely altered with respect to the degenerate case. This can be appreciated in ref.\,\cite{Chang:2018rso} for the case of inelastic fermionic dark matter.
In conclusion, a dedicated analysis is needed to investigate the impact of supernova constraints in our scenario. We leave this for future work. Meantime, we shall comment that these constraints are expected to test masses up to $m_1+m_2\lesssim\mathcal{O}(0.1)$ GeV. 
We have shown that the experiments discussed in sec.\,\ref{sec:exp} attain their strongest sensitivities at larger masses. Therefore, that regions of the parameter space are left untouched by supernova constraints.

\section{Conclusions}
\label{sec:conclusions}

Let us summarize the main results of our work.
We have focused on a dark sector containing a pair of non-degenerate dark scalars with masses in the MeV-GeV range. The lightest one is stable, at least on the timescales relevant for our analysis, and it could play the role of the DM.
We have entertained the possibility that the dark sector communicates with the SM fermions via the effective operators in eq.\,\eqref{eq:EFT}. 
We have studied the sensitivities to this scenario of two fixed-target experiments, CHARM and the SHiP proposal, and two proposed LHC experiments, FASER and MATHUSLA. 
In these accelerator experiments dark scalars are produced from the collision of SM particles. If the interactions in eq.\,\eqref{eq:EFT} are weak and/or the mass splitting between the scalars is small, the heaviest dark particle travels macroscopic distances before decaying and leading to a signal in a far placed detector.
We have compared the projected sensitivities of these experiments with constraints from other probes, in particular searches at LEP, LHC and BaBar.

The operators in eq.\,\eqref{eq:EFT} could be generated for instance by the exchange of a $Z^{\prime}$ boson, coupled both to the visible and the dark sectors, see\,\eqref{eq:heavy_Zp}. Other possibilities exist, as mentioned in sec.\,\ref{sec:intro}.
Motivated by the fact that the mediator can be produced on-shell at the LHC, if light enough, we have investigated this $Z^{\prime}$ model.
We have considered both the case of 
a heavy $Z^\prime$ with a mass above the TeV and of a relatively light $Z^\prime$ (concretely, a dark photon with a mass of $40$ GeV). 

Our main results are shown in figs.\,\ref{fig:Ship},\,\ref{fig:Zp_heavy} and\,\ref{fig:dark_photon}.
As evident, the future experiments that we have studied can significantly improve current constraints.
In particular, we have shown that the sensitivity reach of the MATHUSLA experiment can surpass those from future searches at the HL-LHC. This have been demonstrated for the case of the $Z^{\prime}$ model, for which an appropriate comparison with LHC constraints have been performed. 
The phenomenology explored here is different from that arising in models with light mediators ($\lesssim1$ GeV) since, in the experiments that we have considered, the latter can be directly produced on-shell in decays of mesons, see e.g.\,\cite{Curtin:2018mvb,Feng:2017uoz,SHiP:2020noy,Berlin:2020uwy,Berlin:2018jbm,Berlin:2018bsc,Berlin:2018pwi}.
Therefore our analysis complement other studies for the search of dark sectors.

\acknowledgments

We thank O. \'Eboli and R. Franceschini for useful discussions. EB acknowledges financial support from FAPESP under contracts 2015/25884-4 and 2019/15149-6, and is indebted to the Theoretical Particle Physics and Cosmology group at King's College London for hospitality. M.T. acknowledges support from the INFN grant ``LINDARK,'' the research grant ``The Dark Universe: A Synergic Multimessenger Approach No. 2017X7X85'' funded by MIUR, and the project ``Theoretical Astroparticle Physics (TAsP)'' funded by the INFN.

\appendix
\section{Useful equations for computations in the EFT}\label{app:useful_eqs}

We collect in this appendix the expressions that we have used to compute the production and decay of dark particles. For concreteness, we will write the effective operators in terms of the vector and axial couplings defined in eq.\,\eqref{eq:vector_axial_coupls}. To compute the interactions of light mesons with a pair of dark particles we follow ref.\,\cite{Bishara:2016hek,Bertuzzo:2017lwt}, including the dark current in the Chiral Perturbation Theory Lagrangian. We also compute the Wess-Zumino-Term following ref.\,\cite{Pak:1984bn} to consider interactions involving photons. The relevant interactions are given by 
\be\label{eq:Lmesons}
{\cal L}_{\rm mes} = \frac{J_\phi^\mu}{\Lambda^2}  \left[ \sum_P \left( \tilde{c}_P F\, \partial_\mu P  + \frac{e\, c_P}{8\pi^2\,F} \epsilon_{\alpha\beta\rho\mu} F^{\alpha\beta} \partial^\rho P\right) +i  \sum_{P'} c_{P'} \left( \partial_\mu \bar{P}' P' - \bar{P}' \partial_\mu P' \right)\right] \ ,
\ee
{where $F \simeq 93$ MeV is the pion decay constant, $F_{\alpha\beta}$ is the photon field strength and $J_\phi$ is the dark current defined in eq.\,\eqref{eq:dark_current}. The first sum is taken over the pseudoscalar mesons $P=\left\{\pi^0, \eta_1, \eta_8\right\}$\footnote{We denote by $\eta_1$ the pseudoscalar meson associated with the identity generator and with $\eta_8$ the meson associated with the diagonal $T_8$ generator of $SU(3)$.}, while the second sum is taken over the conjugate pairs $P' = \left\{ \pi^+, K^+, K^0\right\}$ and $\bar{P}' = \left\{ \pi^-, K^-, \bar{K}^0\right\}$, with coefficients 
\be\label{eq:axial_couplings_mesons}
\tilde{c}_\pi = c_{A_d} - c_{A_u} \ , ~~ \tilde{c}_{\eta_8} =-  \frac{c_{A_u} + c_{A_d} - 2 c_{A_s}}{\sqrt{3}} \ , ~~ \tilde{c}_{\eta_1} = - \sqrt{\frac{2}{3}} (c_{A_u} + c_{A_d} + c_{A_s}) \ , 
\ee
\be\label{eq:WZW_coefficients}
c_{\pi^0} = 2 c_{V_u} + c_{V_d}\ , ~~ c_{\eta_8} = \frac{c_{V_d} - 2(c_{V_u} + c_{V_s})}{\sqrt{3}} \ , ~~ c_{\eta_1} = \sqrt{\frac{2}{3}} (c_{V_d} + c_{V_s} - 2 c_{V_u}) \ ,
\ee
and 
\be\label{eq:PP_coefficients}
c_{K^0} = c_{V_s} - c_{V_d} \ , ~~ c_{K^+} = c_{V_s} - c_{V_u}\ , ~~ c_{\pi^+} = c_{V_d} - c_{V_u}\ .
\ee
Notice that the pseudoscalar mesons $\eta_8$ and $\eta_1$ appearing in eq.\,\eqref{eq:Lmesons} are not the mass eigenstates. To rotate to the mass basis we will follow Ref.\,\cite{Escribano:2005qq} and write
\be
\begin{pmatrix}
\eta \\ \eta' 
\end{pmatrix} = \frac{1}{\sqrt{2} F} 
\begin{pmatrix}
f^0_\eta & f^8_\eta \\
f^0_{\eta'} & f^8_{\eta'}
\end{pmatrix}
\begin{pmatrix}
\eta_1 \\ \eta_8
\end{pmatrix} \ ,
\ee
where\,\cite{Escribano:2005qq}
\be
\begin{pmatrix}
f_\eta^8 & f_\eta^1 \\
f_{\eta'}^8 & f_{\eta'}^1
\end{pmatrix} = 
\begin{pmatrix}
f_8 c_8 & - f_0 s_0 \\
f_8 s_8 & f_0 c_0
\end{pmatrix} = \sqrt{2} F
\begin{pmatrix}
1.19 & 0.18 \\
-0.48 & 1.17
\end{pmatrix}\ .
\ee
In the mass basis the couplings of eqs.\,\eqref{eq:axial_couplings_mesons} and\,\eqref{eq:WZW_coefficients} become
\al{\label{eq:coupling_mesons}
\tilde{c}_\pi &= c_{A_d} - c_{A_u} \ ,  & ~~  c_\pi & = 2 c_{V_u} + c_{V_d}\ , \\
\tilde{c}_{\eta} & = -[0.727\,  (c_{A_u} + c_{A_d}) - 0.648\, c_{A_s} ]  \ ,  & ~~  c_\eta & = - (1.45 c_{V_u} - 0.73 \, c_{V_d} + 0.65\, c_{V_s})\ , \\
\tilde{c}_{\eta'}&  = - [ (0.589 \, (c_{A_u} + c_{A_d} ) + 0.799\,  c_{A_s} ]\ , & ~~ c_{\eta ' } & = -(1.18 \, c_{V_u} - 0.59\, c_{V_d} - 0.80 \, c_{V_s})\ .
}
The couplings $\tilde{c}_P$ refer to the case where the dark current couples to an axial fermion current (see eq.\,\ref{eq:EFT2}) while we focus on vector interactions. They are reported here only for sake of completeness.
For the vector mesons we use instead the matrix elements that can be found in the literature. More specifically for the heavy J/$\Psi$ and $\Upsilon$ we follow ref.\,\cite{Yeghiyan:2009xc} and write 
\be\label{eq:V_matr_el}
\langle 0 | \bar{q} \gamma^\mu q | V\rangle = f_V^q M \epsilon_V^\mu \ , ~~~~ 
\langle 0 | \bar{q} \sigma^{\mu\nu} q | V\rangle  = - i f_V^q \left( p^\mu \epsilon_V^\nu(p) - \epsilon_V^\mu(p) p^\nu \right)\ ,
\ee
where $M$ is the vector meson mass. All other matrix elements vanish.
Introducing the Wilson coefficients of eq.\,\eqref{eq:vector_axial_coupls}, we define $f_V=c_{V_q}\, f_V^q $, where $q$ is quark flavor correspondent to the meson $M.$
Since $f_V$ is connected to the quarkonium wave function and it is difficult to compute it from first principles, we will express all our results in terms of the (known) decay widths into electrons, as explained later. 
For the light vector mesons ($\rho$ and $\omega$) we instead use the results of ref.\,\cite{Straub:2015ica}.
Considering the quark composition of the mesons wave functions, we write $f_V=\sum_q c_{V_q}\,f_V^q$, with 
\be\label{eq:f_rho_omega}
f_\rho  = \left( 1.68 \, c_{V_u} - 1.59 \, c_{V_d} \right) F \ , ~~~ 
f_\omega = \left( 1.46 \, c_{V_u} + 1.53 \, c_{V_d} \right) F \ .
\ee
Notice that for the democratic choice of the Wilson coefficients in fig.\,\ref{fig:Decays} ($c_{V_f}=1$) the effective coupling of the $\rho$ meson is suppressed with respect to the analogous one for the $\omega$ meson.
We are now in the position to give explicit formulas for the decay widths.

\subsection{Production of dark particles via mesons decays}
Under the assumption $c_{f_R} = c_{f_L}$ the relevant production channels are {\it (i)} $V \to \phi_1 \phi_2$ with $V = \left\{ \rho, \omega, J/\Psi, \Upsilon\right\}$ the vector mesons, and {\it (ii)} $P \to \phi_1\phi_2 \gamma$ with $P = \left\{ \pi^0 , \eta, \eta'\right\}$ the pseudoscalar mesons. Denoting by $M$ the meson mass and defining
\be
\A{x}{y}{z}= 1 + \frac{\left(x-y\right)^2}{z^2} - 2 \,\frac{x+y}{z}\ ,
\ee
the decay widths are given by
\al{
\Gamma(V \to \phi_1 \phi_2)  &=  \frac{f_V^{2} M^3}{48\pi\,\Lambda^4} \A{m_1^2}{m_2^2 }{M^2} \left[ 1- \frac{(m_1-m_2)^2}{M^2} \right]^{1/2} \left[ 1- \frac{(m_1+m_2)^2}{M^2} \right]^{1/2} \ , \\
\frac{ d\Gamma(P  \to \phi_1\phi_2\gamma)}{ds}  & = \frac{2\,e^2\, c_P^2\,}{3\, (4\pi)^7 \, F^2\, \Lambda^4} \frac{M^3\,\left(M^2 - s\right)^3}{s^2} \left[ 1 + \frac{\left(m_2^2 - s\right)^2}{M^4} - 2 \frac{m_1^2\left(m_2^2 +s \right)}{M^4} \right]^{3/2} \ .
}
In the second equation $e$ is the electric charge, the couplings $c_P$ are defined in eq.\,\eqref{eq:coupling_mesons}, and the total decay width is obtained integrating from $(m_1+m_2)^2$ and $M^2$. As for the first equation, in the case of the $\rho$ and $\omega$ mesons we use eq.\,~\eqref{eq:f_rho_omega} to express $f_V$ in terms of the Wilson coefficients. In the case of the heavy quarkonia vectors $J/\Psi$ and $\Upsilon$ we instead use
\be\label{eq:decay_width}
\Gamma(V \to e^+ e^-) = \frac{f_V^{q\,2}\, e^4\, Q_q^2}{12\,\pi\, M}
\ee
to express $f_V^q$ in terms of the known decay width into an electron-positron pair. The electric charge of the relevant quark is denoted by $Q_q$, respectively 2/3 and -1/3 for the $J/\Psi$ and $\Upsilon$ mesons.

\subsection{Decays of $\phi_2$}
We collect now the equations used to compute the decay widths of $\phi_2$. We define the auxiliary functions
\al{
F_1 & = 2 \left(m_1^2-M^2 \right) \left(m_2^2-M^2 \right)  + 2 s \left(m_1^2 + m_2^2 + 2 M^2 \right) - 6s^2 \\ 
& \qquad{} + 8\, s\, M^2  \left(\frac{(s+M^2 -m_1^2)^2}{4\,s\,M^2}-1\right)^{1/2}\left(\frac{(s+M^2 -m_2^2)^2}{4\,s\,M^2}-1\right)^{1/2} \ , \\
F_2 & = \left(m_1^2 -M^2 \right) \left(m_2^2 -M^2 \right) - m_1^2\,m_2^2\, \A{M^2}{s}{m_1^2}^{1/2} \A{M^2}{ s}{ m_2^2}^{1/2} \\ 
& \qquad {}+ \left( m_1^2+m_2^2+2 M^2 \right) s - 3\,s^2\ ,
}
in terms of which we obtain
\al{\label{eq:phi2dec}
\Gamma(\phi_2 \to \phi_1 V) & = \frac{f_V^2\, m_2^3}{16\pi\, \Lambda^4} \left[ 1- \frac{(m_1 - M)^2}{m_2^2} \right]^{3/2} \left[ 1- \frac{(m_1 + M)^2}{m_2^2} \right]^{3/2}\ , \\
\frac{d\Gamma(\phi_2 \to \phi_1 P\bar{P})}{ds} & = \frac{2\,c_{P'}^2}{3(32\pi)^3 \,\Lambda^4} \frac{ F_1^{\phantom{\,}3} - 8\, F_2^{\phantom{\,}3}}{m_2^3\, s^3}\ ,\\
\frac{d\Gamma(\phi_2 \to \phi_1 f\bar{f}) }{ds} & = \frac{1}{64 \pi^3 \, \Lambda^4} \frac{m_1^2}{m_2} \, \A{s}{ m_f^2}{ m_1^2}^{1/2} \, \A{s}{ m_f^2}{ m_2^2}^{1/2} \times  \\
& \qquad \times \bigg\{ c_{V_f}^2 \frac{(s-m_f^2)^2 (m_1^2+m_2^2-m_f^2-s) - m_1^2\, m_2^2 (s+m_f^2)}{s^2}  \\
& \qquad\qquad {} + c_{A_f}^2 \frac{ (m_1^2+m_2^2) (s+m_f^2) - m_1^2 m_2^2 - (s-m_f^2)^2 }{s} \bigg\}\ , \\
\frac{d\Gamma(\phi_2 \to  \phi_1 \gamma P)}{ds} & = \frac{2\, c_P^2\, e^2}{3\,(4\pi)^7 \, F^2\,\Lambda^4} \frac{M^6 \, m_2^3}{s^2} \, \left(1-\frac{s}{m_2^2}\right)^3\, \A{m_1^2}{ s}{ M^2}^{3/2}\ .
}
The total decay width for the $\phi_2 \to \phi_1 \bar{f} f$ decay is obtained integrating $s$ between $(m_1 + m_f)^2$ and $(m_2 - m_f)^2$, while for the $\phi_2 \to \phi_1 \gamma P$ and $\phi_2 \to \phi_1 P \bar{P}$ decays the integration extends between $(M+m_1)^2$ and $m_2^2$, and between $(m_1+M)^2$ and $(m_2-M)^2$, respectively. The couplings $c_P$ and $c_{P'}$ are defined in eq.\,~\eqref{eq:coupling_mesons} and \eqref{eq:PP_coefficients}, while $e$ is the electric charge. For the decay $\phi_2 \to \phi_1 \bar{f}f$ we show both the contributions generated by $c_{V_f}$ and $c_{A_f}$. For all fermions apart the neutrinos we always consider $c_{A_f} = 0$, leaving only the vector contribution. For the neutrinos, on the contrary, we are considering a V-A current, \textit{i.e.} $c_{A_\nu} = -c_{V_\nu}$.
Notice that the decays $\phi_2\rightarrow\phi_1P\bar{P}$ are not shown in fig.\,\ref{fig:Decays} because the effective couplings involved in these decays vanish for the choice of the Wilson coefficients $c_{V_f}=1$ adopted in that plot.

When the momentum transfer in the decay is sufficiently large, chiral perturbation theory breaks down.
Therefore we adopt the following prescription: for $M_2-M_1>2$ GeV the hadronic decays of $\phi_2$ are computed using perturbative QCD (we consider the processes $\phi_2\rightarrow\phi_1f\bar{f}$, described by the third eq.\,\ref{eq:phi2dec} multiplied by $N_c=3$), while in the opposite regime we use the description in terms of mesons presented above.
Notice however that for momentum transfers $\simeq0.5-2$ GeV both approaches are not appropriate. In this window the hadronic decays are difficult to model. An approach in terms of form factors have been pursued in\,\cite{Winkler:2018qyg,Choudhury:2019tss}.

\section{Decay widths of the $Z'$ boson}\label{app:Zp_decay}

We present the formulas for the decay width of the $Z'$ vector mediator with couplings defined in eq.\,\eqref{eq:vector_axial_coupls_Zp}:
\al{
\Gamma(Z' \to \bar{f}f) & = \frac{N_c}{12\pi} M_{Z'} \left( 1-\frac{4 m_f^2}{M_{Z'}^2}\right)^{1/2} \left[ g_{V_f}^2 \left(1+ \frac{2 m_f^2}{M_{Z'}^2} \right) + g_{A_f}^2 \left(1- \frac{4 m_f^2}{M_{Z'}^2} \right)  \right]\ , \\
\Gamma(Z' \to \phi_1 \phi_2) & = \frac{g_\phi^2}{48\pi} M_{Z'} \left[ 1- \frac{\left( m_1 - m_2 \right)^2}{M_{Z'}^2} \right]^{3/2}\left[ 1- \frac{\left( m_1 + m_2 \right)^2}{M_{Z'}^2} \right]^{3/2}\ ,
}
where $N_c$ is the number of colors. As in the previous section, we show also the contribution of the axial coupling to properly compute the neutrino contribution to the decay width.

\bibliographystyle{JHEP2.bst}
\bibliography{DM_bib}

\end{document}